\documentclass{aa}  

\usepackage{graphicx}
\usepackage{txfonts}

\usepackage{color}

\usepackage[urlcolor=blue]{hyperref}
\newcommand{\ltsim}{\raisebox{-.5ex}{$\;\stackrel{<}{\sim}\;$}}
\newcommand{\gtsim}{\raisebox{-.5ex}{$\;\stackrel{>}{\sim}\;$}}

\newcommand{\Halpha}{\ifmmode {\rm H}\alpha \else H$\alpha$\fi}
\newcommand{\Hbeta}{\ifmmode {\rm H}\beta \else H$\beta$\fi}
\newcommand{\Hgamma}{\ifmmode {\rm H}\gamma \else H$\gamma$\fi}
\newcommand{\Hdelta}{\ifmmode {\rm H}\delta \else H$\delta$\fi}
\newcommand{\Lya}{\ifmmode {\rm Ly}\alpha \else Ly$\alpha$\fi}
\newcommand{\Lyb}{\ifmmode {\rm Ly}\beta \else Ly$\beta$\fi}
\newcommand{\HeI}{\ifmmode {\rm He}\,\textsc{i}\,\lambda5876 \else 
                  He\,\textsc{i}\,$\lambda5876$\fi}
\newcommand{\HeII}{\ifmmode {\rm He}\,\textsc{ii}\,\lambda4686 \else 
                   He\,\textsc{ii}\,$\lambda4686$\fi}

\newcommand{\heii}{He\,\textsc{ii}}

\newcommand{\neiv}{[Ne\,\textsc{iv}]}
\newcommand{\nev}{[Ne\,\textsc{v}]}

\newcommand{\cii}{C\,\textsc{ii}]}
\newcommand{\ciii}{\ifmmode {\rm C}\,\textsc{iii}] \else C\,\textsc{iii}]\fi}
\newcommand{\civ}{\ifmmode {\rm C}\,\textsc{iv} \else C\,\textsc{iv}\fi}

\newcommand{\nii}{[N\,\textsc{ii}]}
\newcommand{\niib}{[N\,\textsc{ii}]\,$\lambda \lambda 6548,84$}

\newcommand{\niv}{N\,\textsc{iv}]}
\newcommand{\nv}{N\,\textsc{v}}

\newcommand{\oii}{[O\,\textsc{ii}]}
\newcommand{\oiii}{[O\,\textsc{iii}]}
\newcommand{\ob}{[O\,\textsc{iii}]\,$\lambda\lambda 4959,5007$}
\newcommand{\oiv}{O\,\textsc{iv}]}

\newcommand{\siiv}{Si\,\textsc{iv}}

\newcommand{\tyi}{\hbox{type~1}}
\newcommand{\tyii}{\hbox{type~2}}
\newcommand{\kms}{\hbox{km$\,$s$^{-1}$}}
\newcommand{\logU}{log $\langle$U$\rangle$}
\begin{document}

   \title{Obscured AGN at $1.5$\,<\,${\it}z$\,<\,$3.0$ from the zCOSMOS-deep Survey
   \thanks{The \tyii~AGN catalog with electronic data is available at the CDS via \hbox{anonymous} ftp to 
    cdsarc.u-strasbg.fr (130.79.128.5) or via http://cdsarc.u-strasbg.fr/viz-bin/qcat?J/A+A/}}

   \subtitle{I. Properties of the emitting gas in the narrow-line region}

   \author{M.~Mignoli\inst{1}
          \and
          A.~Feltre\inst{2,3,4}
          \and
          A.~Bongiorno\inst{5}
          \and
          F.~Calura\inst{1}
          \and
          R.~Gilli\inst{1}
          \and
          C.~Vignali\inst{6,1}
          \and
          G.~Zamorani\inst{1}
          \and
          S.J.~Lilly\inst{7}
          \and
          O.~Le~F\`evre\inst{8}
          \and
          S.~Bardelli\inst{1}
          \and
          M.~Bolzonella\inst{1}
          \and
          R.~Bordoloi\inst{9,10}
          \and
          V.~Le~Brun\inst{8}
          \and
          K.I.~Caputi\inst{11}
          \and
          A.~Cimatti\inst{6,12}
          \and
          C.~Diener\inst{13}
          \and
          B.~Garilli\inst{14}
          \and
          A.M.~Koekemoer\inst{15}
          \and
          C.~Maier\inst{16}
          \and
          V.~Mainieri\inst{17}
          \and
          Y,~Peng\inst{18}
          \and
          E.~P\'erez~Montero\inst{19}
          \and
          J.D.~Silverman\inst{20}
          \and
          E.~Zucca\inst{1}
}

   \institute{
  INAF -- Osservatorio di Astrofisica e Scienza delle Spazio di Bologna, OAS, via Gobetti 93/3, I-40129 Bologna, Italy\\
    \email{marco.mignoli@inaf.it}
         \and
   Sorbonne Universit\'es, UPMC-CNRS, UMR7095, Institut d'Astrophysique de Paris, F-75014 Paris, France      
        \and
   Universit\'e Lyon, Univ. Lyon1, Ens de Lyon, CNRS, Centre de Recherche Astrophysique de Lyon UMR5574, F-69230, Saint-Genis-Laval, France
    \and 
    SISSA, Via Bonomea 265, I-34136 Trieste, Italy
        \and
  INAF -- Osservatorio Astronomico di Roma, I-00040, Monteporzio Catone, Italy
         \and
  Dipartimento di Fisica e Astronomia, Universit\`a degli Studi di Bologna, I-40127 Bologna, Italy
         \and
  Department of Physics, ETH Zurich, Wolfgang-Pauli-Strasse 27, 8093, Zurich, Switzerland
         \and
  Aix Marseille Universit\'e, CNRS, LAM, UMR 7326, F-13388 Marseille, France
         \and
  MIT-Kavli Center for Astrophysics and Space Research, 77 Massachusetts Avenue, Cambridge MA, 02139, USA
         \and
  Department of Physics, North Carolina State University, Raleigh NC 27695
         \and
  Kapteyn Astronomical Institute, University of Groningen, P.O. Box 800, 9700 AV Groningen, The Netherlands
         \and
  INAF -- Osservatorio Astrofisico di Arcetri, Largo E. Fermi 5, I-50125, Firenze, Italy
         \and
  Institute of Astronomy, Madingley Road Cambridge, CB3 0HA, UK
         \and
  INAF -- Istituto di Astrofisica Spaziale e Fisica Cosmica di Milano, via Bassini 15, I-20133 Milan, Italy
         \and
  Space Telescope Science Institute, 3700 San Martin Dr., Baltimore MD, 21218, USA
         \and
  University of Vienna, Department of Astrophysics, Tuerkenschanzstrasse 17, 1180 Vienna, Austria
         \and
  European Southern Observatory, Karl-Schwarzschild-Str. 2, D-85748 Garching bei München, Germany
         \and
  Kavli Institute for Astronomy \& Astrophysics, Peking University, 5 Yiheyuan Road, Beijing 100871, China
         \and
  Instituto de Astrofisica de Andalucia, CSIC, Apartado de correos 3004, E-18080 Granada, Spain
         \and
  Kavli Institute for the Physics and Mathematics of the Universe (WPI) , The University of Tokyo, Kashiwa, Chiba 277-8583, Japan
  }

   \date{Received ; accepted }

 
  \abstract
   {The physics and demographics of high-redshift obscured active galactic nuclei (AGN) is
   still scarcely investigated. New samples of such objects, selected with different
   techniques, can provide useful insights into their physical properties.}
   {With the goal to determine the properties of the gas in the emitting region of \tyii~AGN,
   in particular, the gas metal content, we exploit predictions from photoionization models,    
   including new parameterizations for the distance of gas distribution from the central source
   and internal microturbulence in the emitting clouds, to interpret rest-frame
   UV spectral data.}
   {We selected a sample of 90 obscured (\tyii) AGN with 1.45$\,\leq\,${\it z}$\,\leq\,$3.05
    from the zCOSMOS-deep galaxy sample by \hbox{5$\sigma$ detection} of the high-ionization
    \civ~$\lambda1549$ narrow emission line. This feature in a galaxy
    spectrum is often associated with nuclear activity, and the selection effectiveness has
    also been confirmed by diagnostic diagrams based on utraviolet (UV) emission-line ratios.
    We applied the same selection technique and collected a sample of 102 unobscured (\tyi) AGN. 
    Taking advantage of the large amount of multiband data available
    in the COSMOS field, we investigated the properties of the \civ-selected \tyii~AGN, 
    focusing on their host galaxies, X-ray emission, and UV emission lines.
    Finally, we investigated the physical properties of the ionized gas in the narrow-line region
    (NLR) of this \tyii~AGN sample by combining the analysis of strong UV emission lines with 
    predictions from photoionization models. 
}
   {We find that in order to successfully reproduce the relative intensity of UV emission lines
   of the selected high-z \tyii~AGN, two new ingredients in the photoionization 
   models are fundamental: small inner radii of the NLR ($\approx\,$90 pc for 
   $L_{\rm AGN}=10^{45}{\rm erg\,s^{-1}}$), and the 
   internal dissipative microturbulence of the gas-emitting clouds (with $v_{micr}\approx\,$100~\kms). 
   With these modified models, we compute the gas-phase metallicity of the NLR, and our measurements 
   indicate a statistically significant evolution of the metal content with redshift.
   Finally, we do not observe a strong relationship between
   the NLR gas metallicity and the stellar mass of the host galaxy in our \civ-selected \tyii~AGN sample.
   }
   {}


 \maketitle
%

\section{Introduction} 
   

   There is solid observational evidence
   that the growth of super-massive black holes (SMBHs) at the center
   of galaxies is tightly connected to nuclear stellar activity and the
   formation of spheroids \citep{Richstone98}. The correlation between
   the masses of SMBHs and the velocity dispersions and masses of
   their host stellar spheroids \citep{KormRich95, Mago98, FerMer00, Gebh00}
   suggests that the fueling of the central black hole must be
   linked to the spheroid growth. The star formation in the nuclear region
   of the galaxy is indeed predicted to contribute to the overall bulge stellar population
   \citep{SSBCF99, Cid01}. 
   In addition, the similar cosmic trend of the rates of star formation and
   central black hole accretion, both peaking around redshift $\simeq$2
   with an exponential decline at later times \citep{MadDick14, Silverman08, Franc99}, 
   further supports a coevolution of black holes and their host galaxies.
   Both theoretical and observational studies have
   been undertaken to comprehend the evolutionary link between
   active galactic nuclei (AGN) and their host galaxies, but the underlying mechanisms
   responsible for this coevolution are still far from being fully understood.
   In particular, spectroscopic analyses devoted to studying the properties, 
   such as metal content, geometry, and gas dynamics, of the emitting gas in the narrow-line
   region of the AGN, which extends up to the kiloparsec scale \citep[e.g.,][]{Netzer04},
   are fundamental to better understand the AGN-host coevolution in terms of gas-fueling
   toward the central SMBH and of the impact of gravitational accretion on the gas content
   of the host galaxy. A representative census of AGN over a wide range of cosmic
   time is therefore crucial to shed light on the role of central SMBH in this 
   symbiosis with the host. 

   It is now generally accepted that {\it z}$\sim$2 is a critical epoch for 
   galaxy evolution, as both the global star formation rate (SFR) and the
   AGN activity peak at about this redshift \citep{HopBea06, SDSS06, Delvecchio14}.
   The chemical evolution of galaxies and active nuclei at their center in these crucial
   times therefore provides important clues to understand their build-up processes.
   AGN inhabit galaxies where the central SMBH is accreting the surrounding gas
   during its activity phase. The accreted gas possibly keeps the imprint of the past
   star formation of the bulge. By measuring the gas metallicity 
   in regions surrounding the AGN, it could be possible to obtain
   indirect information on the host star formation history.
   In particular, given the larger spatial extent of the NLR emitting gas
   compared to the broad-line region (BLR), estimating the metallicity using the AGN narrow 
   emission lines could be particularly promising as they are more suited as proxy of
   the host galaxy properties. A fortiori, this holds not only at low redshift, but also
   for the high-redshift and high-luminosity AGN samples because the spatial extent of the
   NLR region is larger in higher luminosity AGN, as confirmed by a strong correlation observed
   between the size of the NLR and the luminosity of the optical [O\,\textsc{iii}]\,
   $\lambda 5007$ emission line \citep[e.g.,][]{Bennert02,Liu13,Sun18,Dempsey18}. 
   
   Chemical abundance indicators for the NLR have mainly been calibrated in the
   optical bands. \citet{Storchi-Bergmann98} have explored two metallicity calibrations
   based on three emission-line ratios, namely \ob/\Hbeta, \niib/\Halpha,
   and \oii $\lambda 3727$/\ob, while \citet{Castro17} have recently proposed a metal abundance
   calibration of the N2O2 parameter, defined as the logarithm of the 
   \nii$\lambda 6584$/\oii $\lambda 3727$ ratio. At $z\gtrsim 2$, current optical spectroscopic 
   studies of AGN probe their redshifted UV emission, which prevents exploiting standard
   metallicity indicators. Therefore, the metal content must be estimated by comparing the
   ratios of strong-UV lines detected in AGN spectra with photoionization models
   \citep[e.g.,][]{Groves04,Nagao06,Feltre16}. Recently, \citet{Dors14} have proposed a new index based
   on the \civ, \ciii,\ and \heii\ lines, that is, ${\rm C43=log[(\civ+\ciii)/\heii],}$ to study the
   metallicity evolution of narrow-line AGN in a wide redshift range ($0<z<4$). Moreover,
   UV-line ratios have been found to be a valuable alternative to the standard BPT \citep{BPT81}
   and \citet{Veilleux87} optical diagnostic diagrams to distinguish between AGN and
   star formation activity at high redshift \citep[e.g.,][]{Feltre16,Nakajima18}.\\
   
   In the local Universe, a strong correlation between the stellar mass and metallicity
   is observed in $z\sim$0.1 star-forming galaxies of the Sloan Digital Survey
   \citep[SDSS;][]{Trem04}. This mass-metallicity relation \citep[MZR;][]{Leq79}
   shows a small dispersion ($\approx$0.1~dex in gas-phase O/H ratio) and is relatively
   steep below 10$^{10.5}{\rm M}_\odot$, but flattens at higher stellar masses.
   Moving at high redshift, the MZR evolves with time in such a way that at a
   given stellar mass, the galaxies had a lower gas metallicity in the past.
   The MZR is usually assessed on samples of star-forming galaxies where AGN are removed
   through the exploitation of optical selection criteria, for instance, BPT diagram, the presence
   of broad lines in their spectra, and X-ray data constraints
   \citep[e.g.,][]{Trem04,Erb06,Maio08}. In the literature, the MZR of AGN, its evolution with
   redshift, and a comparison with that of star-forming galaxies is still poorly explored. 
   Previous studies based on rest-UV spectroscopy of AGN support no evolution of the metallicity
   with redshift \citep[e.g.,][]{Nagao06,Matsuoka09,Dors14}. These studies are limited both in
   terms of line detections, as they are mainly based on line ratios of \ciii, \civ, \heii,\ and
   in some cases \nv, and the number of sources, with samples of a few tens of objects at {\it z}$\sim2$
   and even lower numbers at higher redshift. 

   In this work, we present the largest spectroscopic sample of \tyii~AGN at 1.5$<${\it z}$<$3.0
   drawn from the z-COSMOS deep survey \citep{Lilly07}. We use the
   \civ\ emission line as indicator of nuclear activity and its line profile to divide
   the sample into unobscured (broad, \tyi) and obscured (narrow, \tyii) AGN. We focus on the study
   of the excitation properties of the NLR by comparing the ratios of the emission lines
   that are detected in the spectra of our sources.  We also exploit the broad-band photometry
   available to infer the physical properties (e.g., the stellar mass) of the host galaxies.
   Our main goal is to determine the properties of the gas in the NLR of \tyii~AGN, with
   particular regard to the metallicity (as traced by the gas-phase oxygen abundance,
   expressed in terms of $12+{\rm log(O/H)_{gas}}$). To this aim, we exploit predictions
   from photoionization models including new parameterizations for the distance of gas
   distribution from the central source and the velocity of internally microturbulent clouds. 
   We aim at investigating the evolution with redshift of the gas metal content in the NLR
   and compare it with the properties of the host to study the mass-metallicity relation 
   in \tyii~AGN. This work fits into a wider project that is attempting to exploit the high-ionization 
   narrow emission lines to select and characterize samples of obscured AGN both
   at low \citep[using \oiii~$\lambda5007$;][]{Vignali10} and intermediate redshifts 
   \citep[via \nev~$\lambda3427$ detection;][]{Gilli10,Migno13}.
   
   The paper is structured as follows. In Sect.~\ref{sec:sample} we present the
   spectroscopic survey, the sample selection, and the classification into narrow- and broad-line AGN,
   along with the observed spectral properties. Section~\ref{sec:t2sample} summarizes the
   measurements of rest-UV emission lines and stellar masses of the host galaxies for the
   \tyii~AGN sample. In Sect.~\ref{sec:photomodels} we describe the photoionization models
   of the AGN NLR, and we exploit diagnostic diagrams and emission-line ratios to study the
   nebular properties of the AGN NLR. Finally, in Sect.~\ref{sec:discussion} we discuss 
   the properties of the NLR gas in our \tyii~AGN sample, the evolution with redshift of
   its metallicity, and the mass-metallicity relationship in our \civ-selected \tyii~AGN. 
   The summary and conclusions are presented in Sect.~\ref{sec:summary}.

\section{Sample selection}\label{sec:sample}

\subsection{The zCOSMOS-deep spectroscopic survey}

   The Cosmic Evolution Survey \citep[COSMOS,][]{Scoville07} provided
   high angular resolution and good depth Hubble Space Telescope (HST) imaging with single-orbit \hbox{I-band}
   ACS exposures in an equatorial field of two square degrees \citep{Koeke07},
   along with ground-based images in 15 photometric optical and near-infrared bands
   \citep{Capak07}. The zCOSMOS survey \citep{Lilly07} produced
   spectroscopic redshifts for a large number of galaxies in the 
   COSMOS field using VIMOS, a multislit spectrograph mounted
   on the 8m UT3 of the European Southern Observatory's Very
   large Telescope (ESO VLT). The zCOSMOS project has been designed
   to efficiently use VIMOS by dividing the survey into two components:
   the zCOSMOS-bright is a purely magnitude-limited survey, which
   has collected about 20 000 redshifts for objects
   brighter than \hbox{$I\,$=$\,22.5$} across the full COSMOS field.
   This selection culls galaxies mainly in
   the redshift range \hbox{0.1$\,<\,${\it z}$\,<\,$1.2.}
   The second part, zCOSMOS-deep, has targeted about 10 000 \hbox{$B\,<\,$25} 
   galaxies, selected using color-selection criteria to cover the redshift range
   \hbox{1$\ltsim${\it z}$\ltsim$4.} In this case, only the central 1~deg$^2$
   region of the COSMOS field has been observed, and the spectra of the targeted
   objects have been obtained using the \hbox{R$\sim$200}
   LR-Blue grism, which provides a spectral coverage from 3700 to 6800~\AA.
   The spectra were reduced and calibrated with the VIMOS
   Interactive Pipeline Graphical Interface software \cite[VIPGI,][]{Scodeggio05}, 
   while the redshifts were measured by two different zCOSMOS teams, with the
   support of an interactive package \citep[EZ,][]{Garilli10}, and were then visually
   checked and validated. For more details about the zCOSMOS survey,
   we refer to \citet{Lilly09}.  

\subsection{The \civ-selected sample}

   \begin{figure}
   \centering
   \includegraphics[width=8cm]{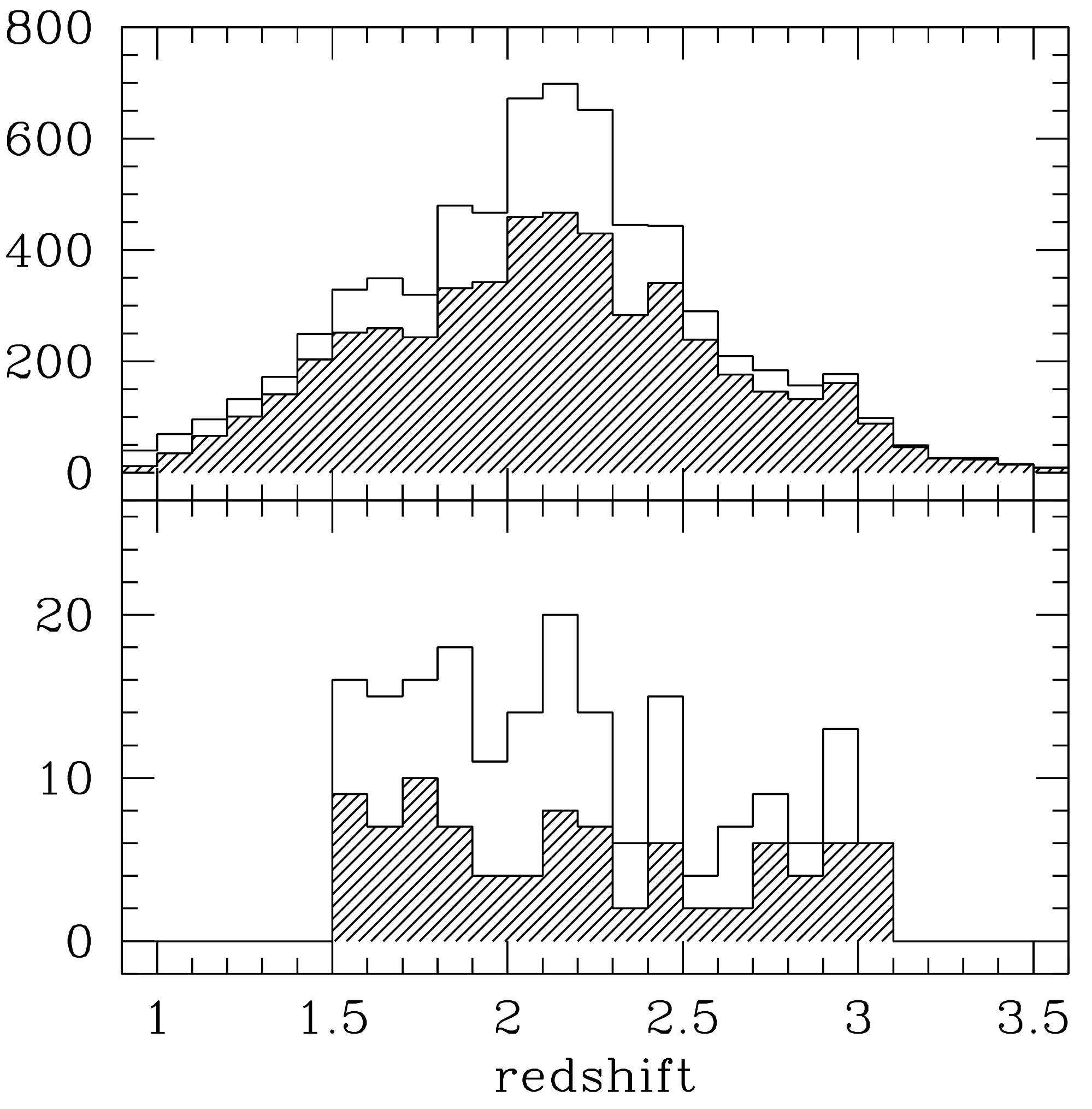}
   \caption{Redshift distributions of zCOSMOS-deep galaxies with $z\ge1$.
   {\it Top panel}: Parent galaxy sample; the hatched histogram shows the galaxies
    with redshift quality flag $\ge\,$1.5. {\it Lower panel}:  \civ-selected AGN sample;
    the hatched histogram shows the \tyii \ sample.}
              \label{Fig1}%
    \end{figure}

  The zCOSMOS-deep survey includes 9523 spectroscopically observed
  objects, but only 7635 of them (80\%) have available redshift measurements.
  A quality flag has been assigned to account for the redshift
  measurement reliability: objects with flags 4 and 3 have very
  secure redshifts, whereas objects with flags 2 and 1 have less secure
  redshifts, with decreasing spectra quality; flag~9 indicates redshifts
  based on a single emission line. 
  Moreover, a decimal place modifier (0.5) is added to denote the agreement
  with the photometric redshift \citep[see][for a detailed definition of the
  confidence classes]{Lilly09}. In this paper we analyze galaxies with flags
  4, 3, 2.5, 1.5, and 9.5 because their redshifts are either secure on their
  own or are confirmed by the associated \hbox{photo-z.} 
  To guarantee that the \civ~$\lambda1549$ emission line is included in
  the observed wavelength range, we limit our analysis to the redshift
  range \hbox{1.45$\,<${\it z}$\,<$3.05}.
  The final analyzed sample contains 4391 galaxies, and their VIMOS spectra
  were visually inspected and measured to select the \civ-emitting objects.
  
  The motivation for using the \civ \ emission line to select high-z AGN
  from the zCOSMOS-deep survey is twofold: on the one hand, the high-ionization
  potential of the triply ionized carbon, C$^{3+}$, makes it a likely signature 
  for nuclear activity, and furthermore, the \civ \ emission line is the most
  intense AGN feature in the UV range, apart from \Lya. On the other hand, 
  in the redshift range culled by the zCOSMOS-deep survey 
  (\hbox{1.0$\,<${\it z}$\,<$3.5}, see the upper panel of Fig.~\ref{Fig1}),
  the \civ \ is the most frequent spectral feature, with more than 85\% of the
  galaxies covering the \civ \ wavelength range in their spectra.
   
  The accurate selection process consists of two main steps. First, the \civ \
  wavelength region is visually inspected in all the galaxy spectra, 
  within the redshift range where the line is visible, to identify possible emitter candidates.
  The complexity of the \civ \ spectral feature did not allow an automatic detection procedure, 
  since the \civ \ line profile could include an AGN emission component, an absorption doublet
  due to the ISM, and a stellar wind feature which should display a P-Cygni shape.
  This preliminary analysis
  reduced the number of galaxies to be further examined to a few hundred,
  allowing a careful inspection of the one- and two-dimensional sky-subtracted
  spectra to eliminate spurious detections (i.e., sky-line and cosmic-ray residuals).
  The final pass of the selection process consists of the measurement, on the
  three-pixel smoothed spectra, of the putative \civ \ emission line using
  IRAF task {\it splot}: we selected as strong emitter all the objects
  with a \civ \ emission peak five times higher than
  the continuum r.m.s. estimated in 50\AA\  windows adjacent to the emission line.
  No \civ-emitter has been found in galaxy spectra with confidence class 1.5,
  probably because the presence of a strong emission line caused the assignment of a
  fair-to-high redshift reliability.
  Following this laborious but accurate selection, we identified 192 
  \civ-selected AGN candidates, and their redshift distribution is shown as
  an empty histogram in the lower panel of Fig.~\ref{Fig1}.

%
 
\subsection{Spectral measurements and AGN classification}\label{sec:AGNclassification}

   \begin{figure}
   \centering
   \includegraphics[width=8cm]{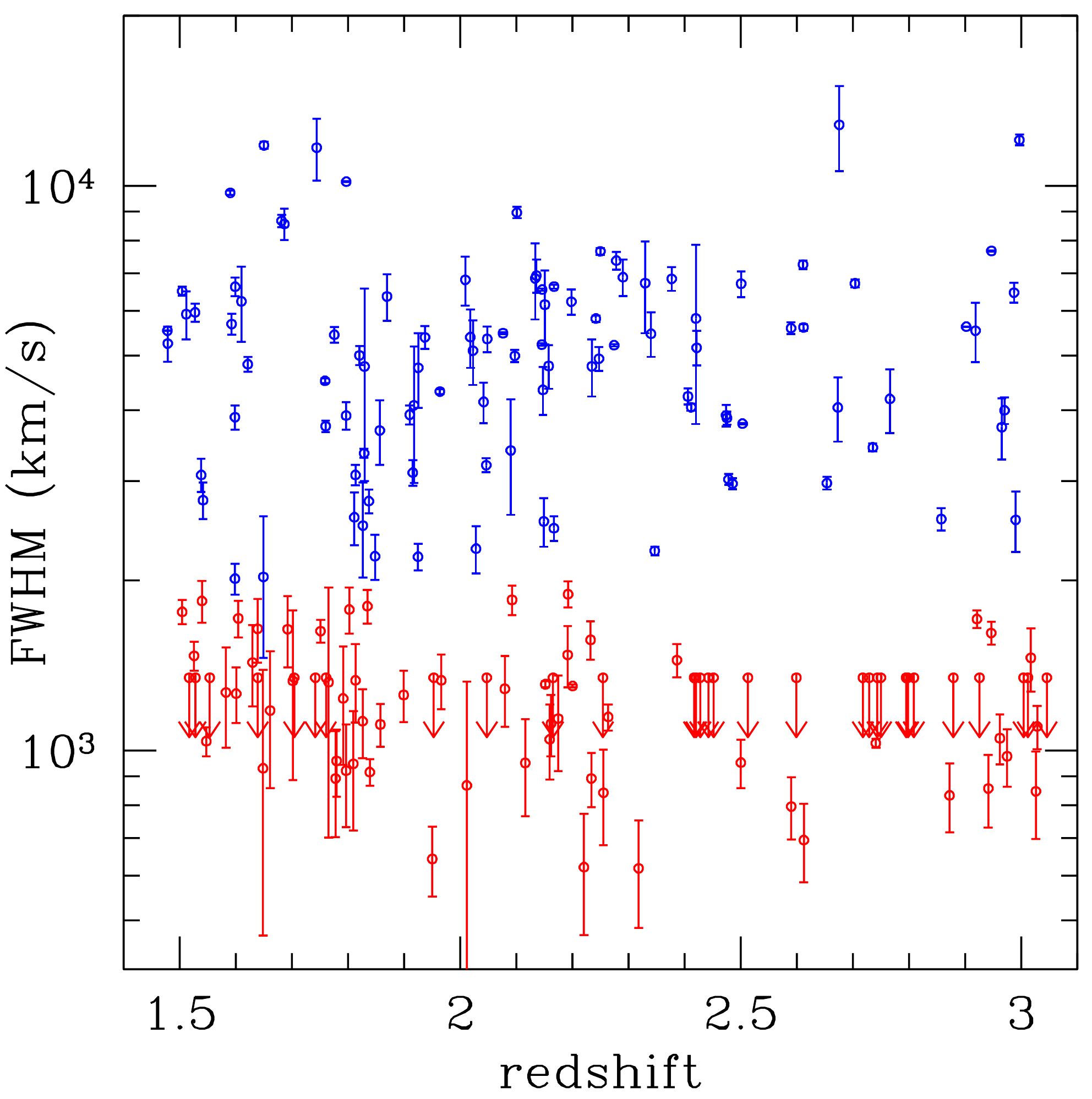}
   \caption{Intrinsic rest-frame FWHMs of the \civ \ emission line in our sample as a function of redshift.
    Objects with line FWHM strictly less than 2000 \kms \ are classified as \tyii\  AGN (red symbols);
    unresolved lines 
    are shown as downward arrows. Broad emission-line objects (with FWHM$>$2000~\kms)
    are classified as \tyi~AGN and \hbox{plotted} with blue symbols.
    }
              \label{Fig2}%
    \end{figure}

The first aim of the spectral line measurement is to classify the selected AGN candidates
into narrow-line (\tyii) and broad-line (\tyi) objects. Because the \civ \ emission line is 
by construction ubiquitous and has an adequate signal-to-noise ratio (S/N), it was 
accurately fit in all spectra, assuming a linear continuum and a Gaussian profile. 
In order to evaluate the quality of the line profile fitting and to estimate the error
in the measurements of the full width at half-maximum (FWHM), we performed four different line fits,
using both the original raw and the three-pixel smoothed spectra, and adopting
different wavelength regions in the continuum level determination. The results of the
first set of spectral measurements are presented in Fig.~\ref{Fig2}, where the intrinsic
FWHM of the \civ \ emission line is shown as a function of redshift. To estimate the
intrinsic width of the line, the observed FWHM was deconvolved by subtracting in quadrature
the instrumental resolution, FWHM$_{inst}$=1350~\kms, as determined from sky lines.
We used the width of the \civ \
line to carry out the classical AGN optical classification: 
the objects were identified as \tyii~AGN if the rest-frame
FWHM was less than 2000~\kms \ (to be conservative, we required that the FWHM 
measurement plus its associated 1$\sigma$ error should be below the threshold value);
otherwise, we classified them as \tyi~AGN. A few objects, especially at lower redshifts,
inhabit a part of the FWHM space near the separation limit, but
an analysis of the spectral and morphological properties of the two AGN samples 
confirmed the goodness of the adopted threshold of 2000~\kms. Out of the 192 \civ-selected
AGN candidates, 90 are classified as \tyii~AGN, and their redshift distribution
is shown in the lower panel of Fig.~\ref{Fig1} as the dashed histogram. 
The remaining 102 selected \tyi~AGN were
already identified as such during the zCOSMOS redshift measurement process,
where broad emission-line objects were flagged, and most importantly, no object
previously identified as a quasi-stellar object (QSO) was missed by our selection process in the redshift 
range of interest.
 
\subsection{Spectral properties of the AGN samples}

   \begin{figure}
   \centering
   \includegraphics[width=8cm]{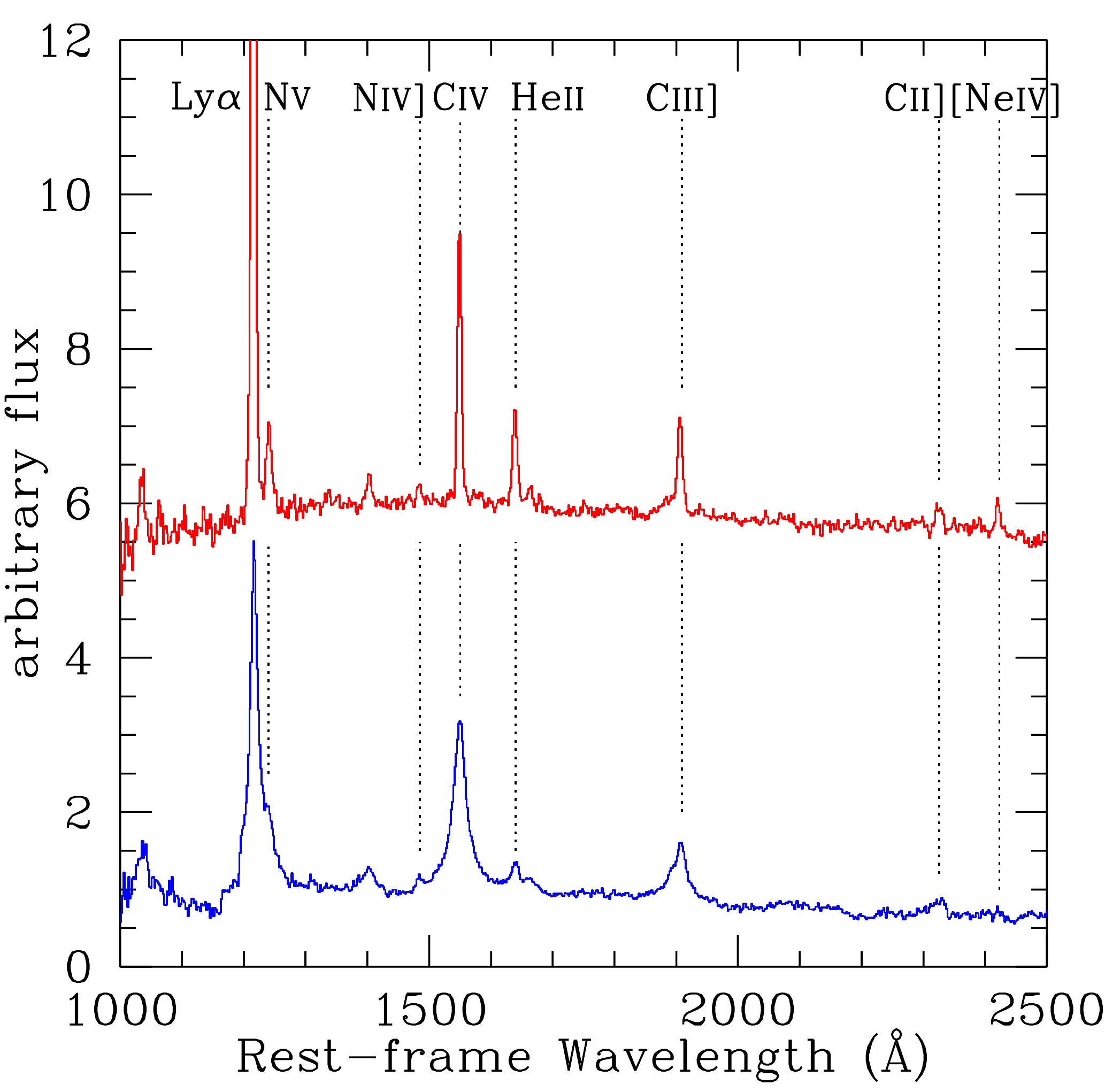}
   \caption{Composite spectra of the two AGN samples with the identification of the main
   emission lines: the lower (in blue) and upper (in red) spectra represent 
   the average spectra of the \civ-selected \tyi\ and \tyii~AGN, respectively.
   The flux normalization is arbitrary, and the spectra are offset for clarity. 
    }
              \label{Fig3}%
    \end{figure}

 We investigated the average spectral properties of the two
 AGN samples by generating and analyzing their composite spectra,
 obtained by coadding all available zCOSMOS spectra included in each class.
 To create the composite, each spectrum was shifted to the rest-frame
 according to its redshift (with a 2.0\AA \ bin width, to match the instrumental pixel
 size at the median redshift of the sample) and normalized to a common
 continuum region, which is always observed in the spectral window. 
 An identical weight was assigned to each
 individual spectrum to avoid biasing the final composite toward the brightest objects. 

 Figure~\ref{Fig3} shows the average spectra corresponding to the \tyi\ (bottom blue curve) 
 and \tyii\ (top red curve) AGN samples. Both spectra show the typical AGN emission lines as labeled in the figure. 
 A summary of the main line parameters is also provided
 in Table~\ref{linepar}. All the emission lines in the composite spectrum of the \tyii~AGN sample
 are consistent with being unresolved (the upper limit of 1350~\kms \ corresponds
 to the spectral resolution of the zCOSMOS-deep data), with the possible exception
 of the \nv~$\lambda$1240 line, which is apparently resolved.
 Comparing the equivalent widths (EW) measured in our \hbox{\tyi~AGN composite} with the
 values reported in \citet{Harris16} from the BOSS composite spectra 
 (third and fifth column of Table~\ref{linepar}, respectively), 
 we observe a significant difference: the larger EWs that are measured in our \civ-selected
 sample can be explained to be a consequence of the Baldwin effect \citep{Bald77} because our AGN
 are fainter than thoses of the BOSS survey.

%
%
   \begin{table}
      \caption[]{Spectral measurements in AGN composite spectra.}
         \label{linepar}
     $
     \resizebox{0.49\textwidth}{!} {         
       \begin{tabular}{lrcrrc}
            \hline
            \noalign{\smallskip}
            &\multicolumn{2}{c}{NL-AGN}   &\multicolumn{2}{c}{BL-AGN}  
            & BOSS-QSO \\
            \noalign{\smallskip}
            line & EW & FWHM  & EW & FWHM  & EW \\
             & [\AA] & [\kms] & [\AA] & [\kms] & [\AA] \\
            \hline
            \noalign{\smallskip}
            \Lya~~~~~{$\lambda$1216} & 155 & $<$1350 & 60: &$\sim$4000 & 85.7 \\
            \nv~~~~~~{$\lambda$1240} & 12.5 &  $\sim$2500 & 15: & $\sim$4000 & 9.2 \\
            \niv~~~~{$\lambda$1485} & 1.9 & $<$1350 & 2.3 &  & 0.15 \\
             \civ~~~~~{$\,\lambda$1549} & 30.6 & $<$1350 &58.0 & $\sim$4200 & 37.4 \\
             \heii~~~~{$\,\lambda$1640} & 12.4 & $<$1350 & 9.0 & $\sim$4000 & 1.5 \\
             \ciii~~~~{$\lambda$1909} & 14.5 & $<$1350 & 30.0 & $\sim$4500 & 25.9 \\
             \cii~~~~~{$\lambda$2326} & 5.1 &  & 7.1 & $\sim$3600 & 0.5 \\
             \neiv~{$\lambda$2423} & 5.5 & $<$1350 & 1.4 &  & 0.9 \\
           \noalign{\smallskip}
            \hline
         \end{tabular}
     }
     $
  \tablefoot{All values are in the source rest-frame. The FWHM of a few low S/N lines is
   highly uncertain, so its value is missing in the table; FWHM upper limits refer to unresolved emission lines.
   Because the \Lya \ and \nv \ lines are heavily blended in the broad-line AGN composite, their measured EWs
   are uncertain. In the last column, EWs from the BOSS Quasar composite \citep{Harris16} are listed. 
   }
   \end{table}
%
 

\subsection{X-ray properties of the AGN samples}\label{sec:X-ray}

The deepest X-ray coverage of the entire zCOSMOS-deep region is granted by
the COSMOS-Legacy survey \citep{Civ16}, a {\it Chandra} program that has
observed the COSMOS field with an effective exposure of $\simeq$160~ks
over the central 1.5 deg$^2$. 
Of the 192 \civ-selected AGN candidates, 147 (77\%) are detected by Chandra
within 2.5” from the optical coordinates. The mean displacement between the AGN
optical positions and their X-ray detected counterparts is 0\farcs46, and all of the
associations are validated by the maximum likelihood technique \citep{Mar16}.
Furthermore, the six sources with a spatial offset larger than 1\arcsec\ are the true X-ray
counterparts of the \civ-selected AGN because the likelihood of finding a spectroscopic 
AGN in an X-ray error box is strongly increased by their low surface densities \citep{Zamo99}.

The X-ray detection rate is significantly different for the two AGN spectral classes:
more than 94\% (96/102) of \tyi~AGN are detected by {\it Chandra}, whereas for the
\tyii~AGN, the fraction drops to 57\% (51/92). This different behavior cannot be ascribed
to a significantly different vignetting-corrected (i.e., effective) exposure time for the two
populations.
\citet{Vignali14} also found a striking difference in the rate of X-ray detections for AGN selected
at z$\sim$1 by the presence of the high-ionization [\ion{Ne}{v}] line: 94\% of the objects with broad lines
in their optical spectra showed a Chandra X-ray counterpart, while the narrow-line AGN are detected 
in only one-third of the cases. To verify that this behavior is not related to the selection criteria,
we investigated the X-ray detection rate in the SDSS: \citet{Paris14}
cross-correlated the Data Release 10 Quasar (DR10Q) catalog with the second XMM-Newton Serendipitous
Source Catalog \citep[Third Data Release, 2XMMi-DR3;][]{Watson09}. Analyzing the DR10Q catalog,
we find that 92.4\% of the SDSS quasars are detected in the X-rays. To perform a similar
analysis for the obscured AGN, we cross-correlated the \tyii~SDSS quasars presented in
\citet{Reyes08} with the third XMM-Newton Serendipitous Source Catalogue
\citep[3XMM-DR5;][]{Rosen16}. Of the 118 \tyii~SDSS quasars included in the XMM-Newton
pointings,   55 match entries in the 3XMM-DR5 catalog, corresponding to an 
X-ray detection rate of 46.6\%.
These results reflect significant obscuration in the X-rays
for the \tyii~AGN (typically, the column density is above 10$^{22}$~cm$^{-2}$)
and reassures us that the adopted FWHM threshold for the spectral classification
effectively separates the \civ-selected~AGN~sample into obscured and unobscured objects.
A more exhaustive analysis of the X-ray properties of the \civ-selected AGN
sample will be presented in a forthcoming paper (Vignali et al., in prep.).

\section{The \civ-selected \tyii~AGN sample}\label{sec:t2sample}

The principal aim of this paper is to investigate the physical properties of NLR emitting
gas and their connection with the properties of the host galaxy. The \tyii~AGN sample
is the ideal laboratory for such studies because both the broad component of the emission
lines and the intense radiation from the nuclear engine are shielded by dust, 
allowing a better analysis of the narrow emission lines and of the spectral energy distribution (SED) of the host galaxy.
In the following, we therefore focus on the properties of the narrow-line AGN and
their host galaxies.

\subsection{UV emission-line measurements}\label{sec:line_measurements}

The selected \tyii~AGN show various emission features in the rest-frame UV spectral range 
(see Fig.~\ref{Fig3} and Table~\ref{linepar}). The analysis of the line ratio diagnostic diagrams,
combined with predictions from photoionization models, provides a powerful tool for probing the physical 
conditions within the narrow-line emitting regions of AGN.

We computed fluxes, velocity dispersions, and EWs of the emission lines by
simultaneously fitting Gaussian profiles to each set of lines, assuming the same FWHM
for all the lines and allowing a small shift of the central positions with respect to the systemic 
redshift in order to account for the well-known offset between emission-line peaks in quasars
\citep{SDSSshift02}. Fixing the same velocity dispersion for all the profiles improves the fit
for low S/N lines. This is indeed reasonable because we assumed that all lines originate from 
the same emitting region (i.e., the NLR), and most of them are dominated by the instrumental point spread function (PSF).
The continuum level was fit, in regions free of known emissions, with a polynomial function. 
We used customized fitting procedures, adapted from the IRAF task {\it splot}, to measure 
seven emission lines: \nv$\lambda1240$ (hereinafter \nv), \niv$\lambda1486$ (\niv), \civ$\lambda1549$ (\civ),
\heii$\lambda1640$ (\heii), \ciii$\lambda1909$ (\ciii), \cii$\lambda2326$ (\cii), and \neiv$\lambda2423$ (\neiv).
The \Lya \ emission line was excluded from the simultaneous fit because it could be severely affected by
resonant scattering in the interstellar and circumgalactic medium, internal kinematics, and emissivity distribution,
which would make the comparison with the outcome of the photoionization models difficult.
We also avoided the line blend \siiv/\oiv \ at $\sim$1400\AA.

Errors on the flux measurements were estimated following the recipe described by \citet{LenzA92}:
\begin{equation}
\sigma(Flux)=\frac{Flux_{TOT}}{(S/N)_{cont}}\times\frac{(FWHM/\Delta\lambda)^{-1/2}}{C_L},
\end{equation}
\noindent
where $Flux_{TOT}$ is the measured flux of the emission line, $(S/N)_{cont}$ is the signal-to-noise ratio
of the continuum close to the line, $\Delta\lambda$ is the spectral sampling, and $C_L$ is the coefficient
computed by \citet{LenzA92} (see their Table~1). 
The computed relative flux errors are $\approx$10\% for the brightest emission lines 
(flux$\,>5\times10^{-17}{\rm erg\,cm^{-2}\,s^{-1}}$),
but they increase up to $\approx$30-40\% for the faintest ones (with 
flux$\,\approx$$2\div5\times10^{-18}{\rm erg\,cm^{-2}\,s^{-1}})$.

\subsection{Stellar masses of the host galaxies}\label{sec:masses}

   \begin{figure}
   \centering
   \includegraphics[width=8cm]{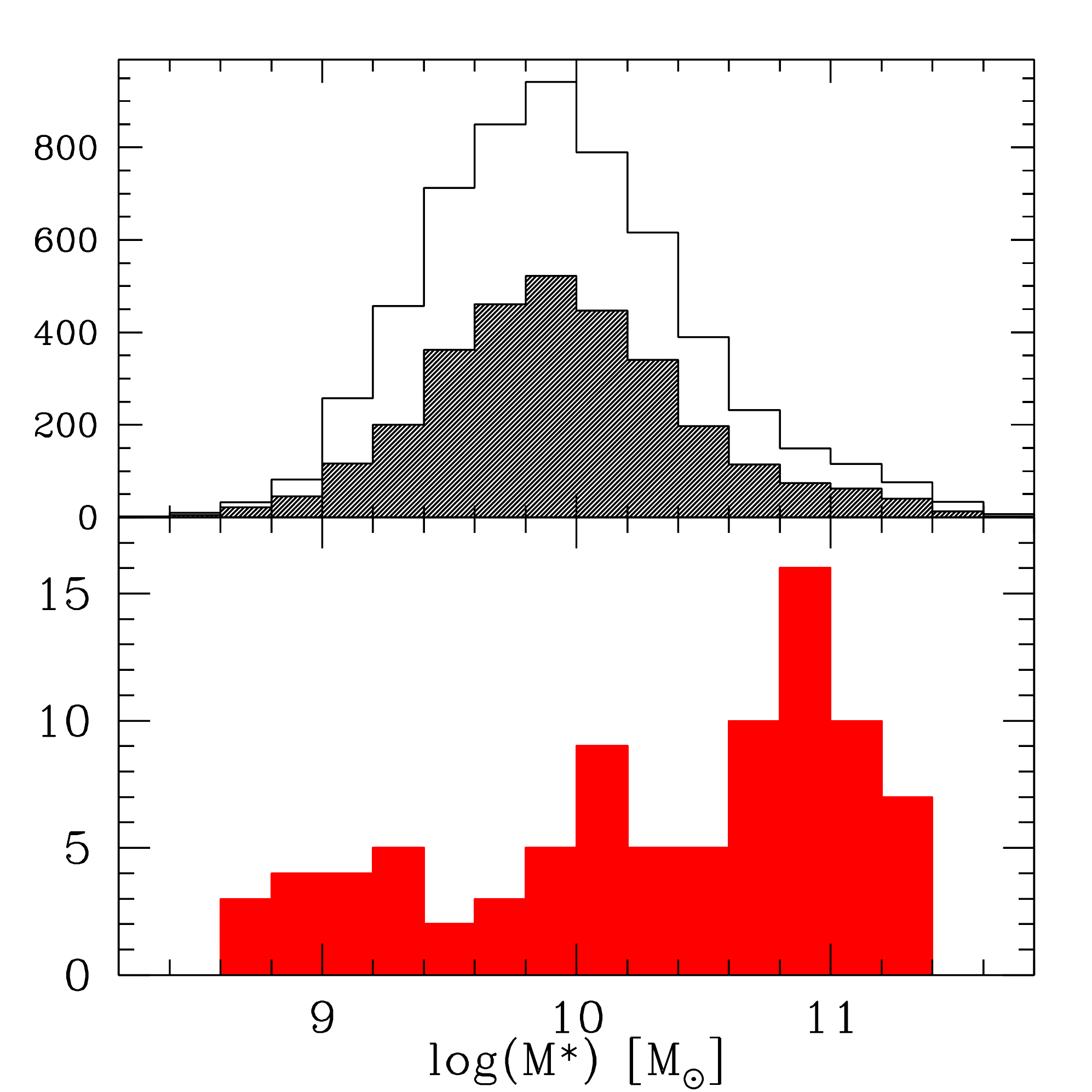}
   \caption{Stellar mass distribution of zCOSMOS-Deep galaxies with 1.45$\,<\,${\it z}$\,<\,$3.05.
   {\it Top panel}: Parent galaxy sample; the hatched histogram shows the galaxies with
   redshift quality flag $\ge2$. {\it Lower panel}: Stellar mass \hbox{distribution} of the galaxies
   hosting the \civ-selected \tyii \ AGN. For these objects the masses are derived from
    a two-component (AGN and galaxy templates) SED fitting. 
    }
              \label{Fig4}%
    \end{figure}
Stellar masses were computed using a two-component (AGN and galaxy) SED fitting technique.
Briefly, we used a combination of AGN and host-galaxy templates to fit the large set of optical
and near-infrared photometry available in the COSMOS field, using a $\chi^2$ minimization
to find the best solution that reproduced the observed flux of each object at a redshift fixed to
the spectroscopic one. In particular, for the AGN component, we adopted the \citet{SDSS06b} 
mean QSO SED as derived from the study of 259 IR-selected quasars with both SDSS and 
{\it Spitzer} photometry, while the galaxy component was described using a library
of synthetic spectra generated from the stellar population synthesis model of 
\citet{BC03}, with different star formation histories and ages and assuming a universal initial
mass function (IMF) from \citet{Chabrier03}. A more detailed description of the method 
can be found in \citet{Bongiorno12}.
The lower panel of Fig.~\ref{Fig4} shows the stellar mass distribution of the host galaxies for the
\tyii~AGN sample. We computed stellar masses for all but two of the \civ-selected \tyii~AGN because the photometric data for two objects did not allow a reliable estimate: nearby objects 
seriously contaminate the data. The host galaxy stellar masses are in the range 
$8.5\times 10^{8}-\;2.2\times 10^{11}{\rm M}_\odot$, with an average (median) stellar mass of 
$2.1\times 10^{10}{\rm M}_\odot$ ($3.6\times 10^{10}{\rm M}_\odot$).
The upper panel of Fig.~\ref{Fig4} shows the mass distribution of the parent sample of
zCOSMOS-Deep galaxies in the redshift range where the \civ \ line is present in the spectral
coverage (empty histogram), along with that of the galaxies with spectroscopic redshift
flag $\ge 2$, represented by the shaded histogram. For the parent sample the stellar masses
were computed following the recipe adopted by \citet{Bolzonella10}. We note that for our
\tyii~AGN sample, the stellar masses estimated by the two methods differ by less than 30\%
for the majority of the cases (85\%), and for the galaxies with the largest mass disagreement,
the higher value is often computed by the method that does not include the AGN contribution.
Even when we take the different calculation  methods into account, the two mass distributions are
significantly different, with the \tyii~AGN host galaxies peaking toward the high-mass tail. 
We used a two-population Kolmogorov-Smirnov (K-S) test to assess the significance
of the difference between the stellar masses of the galaxy parent sample and the \tyii~AGN hosts. We
find that they differ at very high significance ($>6\sigma$).
 
In order to parameterize the relative contribution of the AGN emission over the host galaxy light,
we computed the dereddened luminosity of the two components in the rest-frame K band,
using two distinct values (one for the galaxy and one for the active nucleus) of the dust attenuation
provided by the SED fitting routine.
We defined the ratio between the AGN K-luminosity and that of the host galaxy as "AGN dominance''.
Reassuringly, this ratio correlates with the rest-frame EW of the \civ \ emission line, which is
also a good indicator of the AGN prominence because the EW is the ratio between the flux of the
AGN-powered emission line and the UV stellar continuum (the power-law continuum
from the nuclear engine is largely shielded by dust in obscured AGN). 
Only five of the \civ-selected objects show an AGN contribution lower than 20\%, 
and all of them also belong to the low tail of the \civ \ EW distribution; their rest-frame
equivalent widths are smaller than 10\AA. Because these objects are also undetected in the X-rays,
they are the prime suspects for being low-luminosity AGN, if not actually pure star-forming galaxies.

\section{Comparison with photoionization models}\label{sec:photomodels}

We compared observations with predictions from photoionization models (described in Sect.~\ref{sec:models}) 
to improve our understanding of the excitation properties  of the narrow-line emitting gas in AGN. 
We first explore diagnostic diagrams (Sect.~\ref{sec:diagnostics}) based on UV emission-line ratios and 
these diagrams confirm the efficacy of the \civ \ line to select AGN.
We then perform a simple spectral fitting to estimate the metallicity, expressed in terms of 
the gas-phase oxygen abundance, of the emitting gas in our \civ-selected AGN (Sect.~\ref{sec:line fitting}).

\subsection{Photoionization models}\label{sec:models}

The nebular emission from the NLR of AGN was modeled with single ionization-bounded 
gas clouds using the approach described in \citet{Feltre16}. 
The AGN ionizing spectrum, described as a series of broken power laws \citep[Eq. 5 in][]{Feltre16}, 
is combined with the photoionization code CLOUDY \citep[version c13.03, last described in][]{Ferland13}. 
We included dust and radiation pressure and adopted an open geometry appropriate for gas with a 
small covering factor, that is, the radiation from the illuminated face of the cloud toward the source 
of continuum radiation is allowed to escape without further interacting with the gas.

The models are parameterized in terms of
\begin{itemize}
\item the UV spectral index, $\alpha$, of the incident radiation field ($F_{\nu}\propto \nu^{\alpha}$) 
at the wavelength range of the ionizing photons;
\item the volume-averaged ionization parameter $\langle U \rangle$\footnote{
The models here are labeled in terms of the volume-averaged ionization parameter 
(see equation B.6 of \citealt{Panuzzo03}) instead of the ionization parameter at the edge 
of the Str\"omgren sphere $U_{\rm S}$ (equation 4 of \citealt{Feltre16}).
This parametrization translates into $\langle U \rangle$=9/4$\,U_{\rm S}$. 
Different choices for model-labelling have no impact on the actual CLOUDY computations, 
for which the input parameter is the rate of ionizing photons.}, 
defined as the dimensionless ratio of the number density of H-ionizing photons to that of hydrogen;
\item the hydrogen gas density of the clouds, $n_{\rm H}$;
\item the interstellar (gas+dust phase) metallicity, $Z$;
\item the dust-to-heavy element mass ratio, $\xi_{\rm d}$;
\item the mean distance from the central source of the illuminated face of the NLR gas
distribution, $r_{\rm in}$, hereafter inner radius of the NLR for conciseness;

\item the internal microturbulence velocity of the gas cloud, $v_{\rm {micr}}$.
\end{itemize}

 The values of these parameters that we adopted in our calculations are summarized in Table~\ref{table:modelpar}. 
 Exploring different values for the NLR inner radius, that is, the distance from the center, means
 testing different scaling relations, where the AGN luminosity is linked to the inner radius of the NLR, 
 $r_{\rm in}$, through the scaling relation $L_{\rm AGN}/r_{\rm in}^2 = 10^p \,{\rm erg \,s^{-1}\, cm^{-2}}$. 
 For a given AGN luminosity, the value of the index $p$ depends only on the assumed $r_{\rm in}$ value. 
 We explored two values of $r_{\rm in}$,  90 and 300 pc, and scaled the models to an AGN accretion-disk
 luminosity, integrated over the wavelength range between 0.001 and 10 $\mu$m \citep[Eq. 5 in][]{Feltre16}, 
 of $L_{\rm AGN}=10^{45}\,{\rm erg\,s^{-1}}$.\linebreak
 To compute the transmission of the radiation through the gas cloud, we set the outer boundary of the 
 photoionized gas distribution by stopping our calculations when the electron density fell 
 below 1\% of the number density of atoms of neutral hydrogen or the temperature fell below 100~K.
 The NLR emission extent of our models covers a wide range of values; the outer boundary of 68\% of the models
 lies between $\sim$0.8~pc and $\sim$12~kpc, and the median value
 is $\sim$3.8~kpc. The median luminosity of the \oiii$\lambda5007$ lines in our models is
 $~ 3.6 \times 10^{42} \, {\rm erg\, s^{-1}}$, in agreement with the empirical values shown
 in Fig. 15 of \cite{Storchi-Bergmann18}.

%
%
   \begin{table}
      \caption[]{Main adjustable parameters of the photoionization models}
         \label{table:modelpar}
     \resizebox{0.49\textwidth}{!} {         
       \begin{tabular}{cc}
            \hline
            \noalign{\smallskip}
            Parameter & Adopted Values \\
            \hline
            \noalign{\smallskip}
                Ionizing spectrum & $\alpha= -1.2, -1.4, -1.7, -2.0 $\\
                $\log ( \, \langle U \rangle\, )$ & $-0.65, -1.15,  -1.65 , -2.15, -2.65, -3.15, -3.65, -4.15$ \\
                 $\log(n_{\rm H}/{\rm cm}^{-3})$ & 2.0, 3.0, 4.0 \\  
                  & 0.0001, 0.0002, 0.0005, 0.001, 0.002, \\
                 $Z$& 0.004, 0.006, 0.008, 0.014, 0.01774,  \\
                 & 0.03, 0.04, 0.05, 0.06, 0.07\\
                 $\xi_{\rm d}$ & 0.1, 0.3, 0.5 \\
                 Inner radius & $r_{\rm in }= 90, 300$ pc\\
                 Microturbulence velocity & $v_{\rm {micr}}=0, 100 $ km/s\\
                
           \noalign{\smallskip}
            \hline
         \end{tabular}
         }
   \end{table}

 \subsection{Microturbulent clouds and inner radius of the NLR gas}\label{sec:new_param}
 
 It is worth noting that the AGN NLR model grid adopted in this work is an improvement 
 over \citet{Feltre16} and includes two new adjustable parameters: the inner radius of the NLR, 
 $r_{\rm in }$, and the internal microturbulence velocity of the gas cloud, $v_{\rm {micr}}$.
 These additional terms have been introduced to explain the observed luminosity of the \nv\ compared to 
 \civ\ and \heii\ emission lines, which is commonly underestimated by photoionization models of 
 both broad- and narrow-line emitting regions of AGN \citep{Davidson79}.
 
 To solve the problem of high \nv/\civ\ and \nv/\heii\ ratios observed in AGN, some authors 
 appeal to super-solar metallicities  \citep[e.g.,][]{HamannFerland93, ShemmerNetzer02, Nagao06},
 up to 10 times solar \cite[see also Sec. 7.1.2 of][]{Netzer13}. 
 However, these same high-metallicity models 
 do not reproduce other ratios observed in the spectra 
 of the same sources \citep{Nagao06,Feltre16}, such as \civ/\heii\ and \ciii/\heii,
 which instead favored lower metal abundances.
Other solutions have been proposed 
in the literature, such as "selectively" enhanced nitrogen abundance \citep{HamannFerland92,HamannFerland93}
 and internally microturbulent clouds \citep{Bottorff00,Kraemer07}.
We did not consider models with an overabundance of nitrogen for two reasons:
first, photoionization models agree well with the observed emission from other nitrogen ions, 
such as \niv\ (observed in this sample), without requiring 
an enhanced fraction of nitrogen or highly super-solar metallicity. 
Second, we have computed models in which the nitrogen abundance was increased by 0.15 dex compared 
to that used in \citet{Feltre16} and found no significant improvement in reproducing the 
observed line ratios involving \nv. 
We note that our models already account for the secondary nucleosynthesis production of nitrogen
\citep[equation 11 of][]{Gutkin16} that results from the CNO cycle of stars that are already enriched in carbon
and oxygen and are therefore important at high metallicities \citep[e.g.,][]{Cowley95, Chiappini03}.

As an alternative to exotic metal abundances, we computed models with a dissipative 
turbulence ($v_{\rm {micr}}=100$~\kms) internal to the clouds that converts the 
turbulent motion into heat, following the approach of \citet{Kraemer07}. 
The consequent increase of the electron temperature affects the emissivity of some 
emission lines, and in particular, that of the collisionally excited lines. 
This hypothesis is further supported because the FWHM of the \nv\ line is wider than other high-ionization emission lines that are observed in the \tyii~composite spectrum
(see Table~\ref{linepar}), although the poor resolution of the VIMOS spectra, the partial blending 
with the \Lya, and the doublet nature of the N\,\textsc{v} line mean that 
the line width difference cannot be used to constrain the microturbulence models.
We also explored different scaling relations between the AGN luminosity and the distance
of the inner face of the NLR gas clouds from the central source, that is, models with different values
of the inner radius of the NLR emitting gas ($r_{\rm in} = 90, 300$ pc) for $L_{\rm AGN}=10^{45}\,{\rm erg\,s^{-1}}$. 
A lower inner radius at fixed AGN luminosity translates into an increase in radiation pressure 
at the inner face of the gas cloud $P_{\rm in}$ ($L/4\pi c r_{\rm in}^2 = P_{\rm in}$, 
see also equation 1 of \citealt{Dempsey18}). 


   \begin{figure*}
   \centering
   \includegraphics[width=11cm]{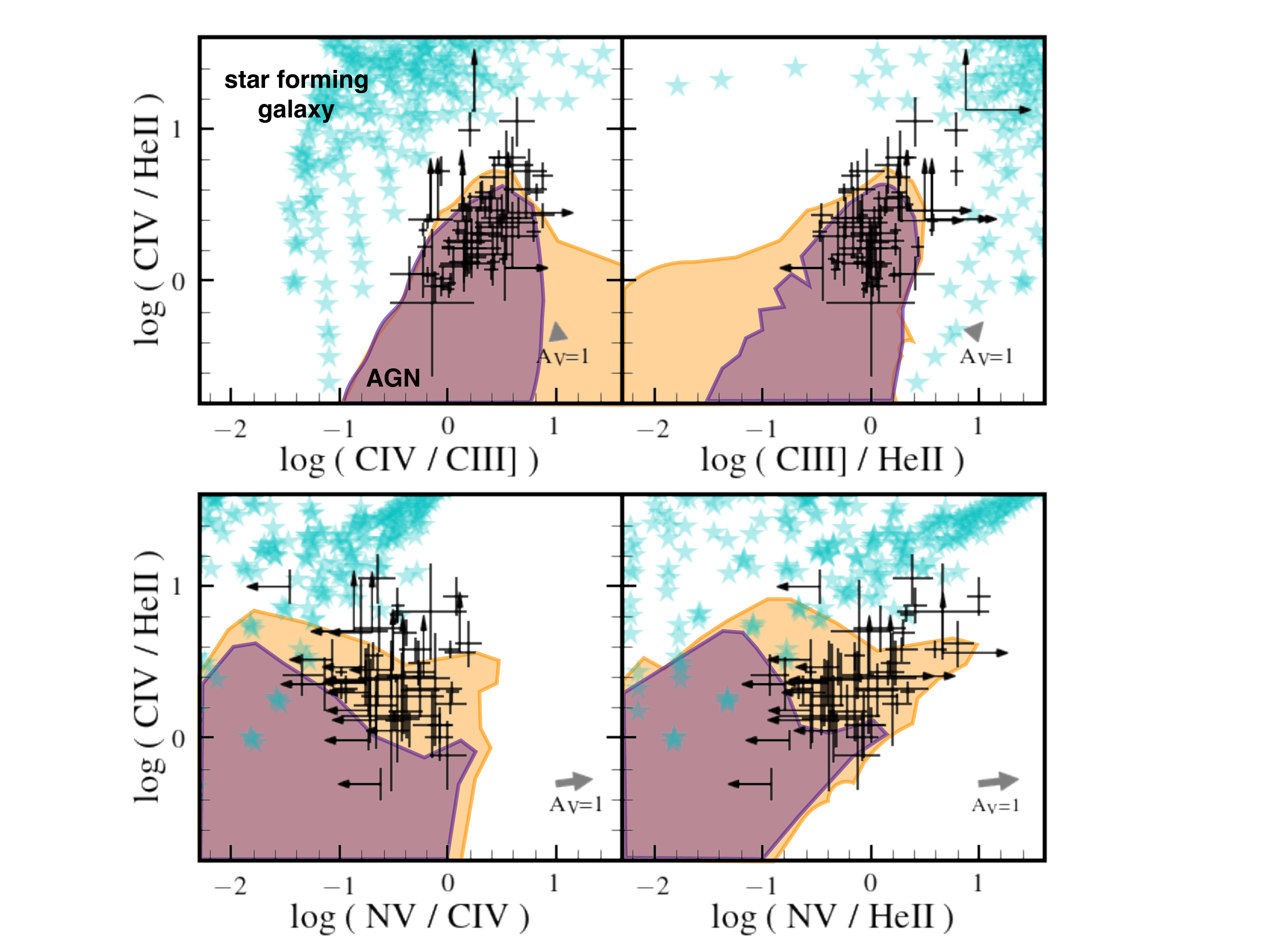}
   \caption{Predictions of the AGN NLR models described in Sect.~\ref{sec:models} in the \civ/\heii\ 
   vs. \civ/\ciii\ (top left), \ciii/\heii\ (top right), \nv/\civ\ (bottom left) and \nv/\heii\ (bottom right) 
   diagrams. Each panel shows data measurements with error bars and upper limits (black symbols)
   of the \civ-selected \tyii~AGN, described in Sect.~\ref{sec:t2sample}, and AGN models (shaded areas) 
   encompassing wide ranges in power-law index, $\alpha$, ionization parameter, $U_{\rm S}$, gas metallicity, Z,
   gas density, $n_{\rm H}$, and dust-to-metal mass ratio, $\xi_{\rm d}$. The purple and orange
   shaded areas indicate AGN models with no microturbulence velocity and an inner radius of 300~pc
   and models with microturbulence velocity $v_{\rm micr}=100$~\kms \ and an inner radius of 90~pc, respectively.
   Cyan stars show the predictions from photoionization models of star-forming galaxies of \citet{Gutkin16}.
   In each panel, the gray arrows indicate the effect of correcting the AGN data by dust attenuation for a galactic attenuation curve \citep{Cardelli89} and $A_{\rm V}$ = 1 mag.}
              \label{Fig5}%
    \end{figure*}

\subsection{Diagnostic diagrams}\label{sec:diagnostics}

We compared the emission-line measurements (Sect.~\ref{sec:line_measurements}) of the \civ-selected
\tyii~AGN with predictions from the photoionization models described above. Figure~\ref{Fig5} shows
four diagnostic diagrams based on UV emission-line ratios, which we found to be good diagnostics of nuclear
activity versus star formation \citep[see Sect. 4 of][and references therein]{Feltre16}.
The \civ/\heii\ versus \civ/\ciii\ and \ciii/\heii\ diagrams, 
originally investigated by \citet{VillarMartin97}, are shown in the two top panels of Fig.~\ref{Fig5}
(left and right, respectively), while the two bottom panels show 
\civ/\heii\ versus \nv/\civ\ and \nv/\heii\ (left and right, respectively).
These line ratios are used in the literature as diagnostics of the ionizing source and to
investigate the gas metallicity of type 2 AGN and {\it z}$\sim$2.5 radio galaxies
\citep[e.g.,][]{DeBreuck00,Vernet01,Groves04,Nagao06,Humphrey08}.

The purple and orange shaded areas in each panel of Fig.~\ref{Fig5} indicate AGN photoionization
models encompassing wide ranges of model parameters: power-law index, $\alpha$, ionization parameter,
$\langle U \rangle$, gas metallicity, Z, hydrogen gas density $n_{\rm H}$ , and dust-to-metal mass ratio
$\xi_{\rm d}$ , as summarized in Table~\ref{table:modelpar}. 
Models with no internal microturbulence and an inner radius of 300 pc as in \citet{Feltre16} are shown
in purple, while models with internal microturbulence ($v_{\rm micr}=100$~\kms) and smaller inner
radius ($r_{\rm in}=90$ pc) are plotted in orange.
For comparison, the cyan stars indicate models of nebular emission from star-forming galaxies by
\citet{Gutkin16}. AGN models, in particular those with internally microturbulent clouds and a small
inner radius, predict line ratios similar to those measured in the spectra of our \civ-selected sample.
Reducing the inner radius and adding the microturbulence has a similar impact on the predictions 
of emission-line ratios and, specifically, lead to the increase in \civ/\heii\ and \nv/\heii\ ratios.

Only a few \civ-selected objects do not lie in the area occupied by pure AGN models in the
diagnostic diagrams; in particular, four objects appear to be outliers with respect to pure
AGN models in all the diagnostic diagrams. 
Two of the four outliers show SEDs with an AGN contribution to the rest-frame K band (see Sect. 3.2)
that is lower than 10\% and very low \civ\ EWs ($<$10~\AA), suggesting a non-negligible contribution
from star-formation activity to the emission lines of these sources. This suggestion is supported by a blue continuum with strong absorption lines in their VIMOS spectra. The other
two outliers present noisy spectra with a possible contribution from the broad component to the \ciii\
emission line. This contribution might artificially increase the \ciii/\heii\ line ratio.
The moderate S/N of the VIMOS spectra did not allow us
to perform a multiple fitting (broad, narrow, and stellar components) of the emission lines,
but we are confident that both the \civ-selection and the removal of the broad-line objects
through FWHM thresholding provided us with a relatively clean sample of narrow-line AGN. 
We also studied diagnostic diagrams based on the EW of \ciii\ and
\civ\ lines, namely EW(\ciii) versus \ciii/\heii\ (EW$-$C3) and EW(\civ) versus \civ/\heii\ (EW$-$C4). 
None of the sources is classified as a star-forming galaxy when the separation criteria of
\citet{Hirschmann19} are adopted. Of the 64 objects that are covered by the \ciii\ line, 8
lie in the composite region (i.e., intermediate between pure AGN and pure star-forming galaxy) 
of the EW$-$C3 diagram, while 15 of the \civ-selected AGN lie
in the composite region of EW$-$C4 \citep[see table 1 and figure 5 of][]{Hirschmann19}.
All the sources except for one are classified as AGN following the separation criteria of \citet{Nakajima18}.
This object is one of the two outliers that show a significant contribution from a stellar 
component in the spectra.
That the diagnostic diagrams contain only a few outliers suggests that contaminations from the BLR and/or
from the star formation (the two main sources of uncertainty in the analysis of the 
UV emission lines) probably do not significantly affect the measurements of the narrow-line fluxes.

\subsection{Physical properties of the nebular gas}\label{sec:line fitting}

To study the excitation properties of our \civ-selected \tyii\ sample,
we compared predictions from the photoionization models (Sect.~\ref{sec:photomodels})
to the observations by fitting emission-line ratios.
In particular, we considered ratios of any possible combination of the emission lines measured in
the spectra of our \civ-selected \tyii~AGN, namely \nv, \niv, \heii, \oiii, \civ, \ciii, \cii ,\ and \neiv.
We adopted a galactic attenuation curve \citep{Cardelli89} for consistency with the fitting procedure
that was used to derive the stellar masses of the host galaxy (Sect.~\ref{sec:masses}).

The \civ-selected \tyii\ sample is ideal to study the excitation properties of the AGN NLR. 
The simultaneous presence of two or more emission lines of ions such as oxygen, nitrogen,
and/or carbon in the same spectra, along with the plethora of the other UV lines available,
provides strong constraints on the photoionization models. 
We computed the likelihood of each model given the data by adopting a simple Bayesian approach,
similarly to \citet[][see their eq. 3]{Vidal-Garcia17}, where our sets of observables are emission-line
ratios instead of spectral indices. We then computed the median values of the posterior distribution functions
and corresponding errors (computed using the 16th and 84th percentile) of the 
parameters of the models (see Sect. \ref{sec:photomodels} and Table \ref{table:modelpar}). 

The line fluxes of our \civ-selected AGN are compatible with a wide range of the volume-averaged
ionization parameter, $-2.8 < $\, \logU \,$< -0.65$, and interstellar metallicities,
$0.002 \lesssim Z \lesssim 0.05$, ($0.1 \lesssim Z/Z_{\odot} \lesssim 3.3$), with median
values of -1.65 and 0.008 ($\sim$half solar), respectively. The present-day solar
(photospheric) metallicity adopted in the models is $Z_{\odot}$=\,$0.01524$ \citep{Bressan12}.

Moreover, the median UV spectral index of the ionizing radiation field for our \tyii~AGN sample
is $\alpha=-1.63$; here, 51\% of the objects with a the median value of the posterior 
distribution function lie in the range $-1.7 < \alpha < -1.55$. We find no preference among the different values of hydrogen gas density, $n_{\rm H}$ and dust-to-metal mass ratio,
$\xi_{\rm d}$. Determining whether these loose constraints are due to a lack of observables connected
with these parameters or to the need of a finer sampling in the models is beyond the scope of this paper.
We also find that about 85\% and 70\% of our sample favor models 
with an inner radius of 90.0~pc for $L_{\rm AGN}=10^{45}\,{\rm erg \,s^{-1}}$
and an internal microturbulence velocity of 100~\kms, respectively.

From the spectral fitting we finally derive values of the gas-phase oxygen abundance,
expressed as $12+{\rm log(O/H)_{gas}}$, of between $7.6< {\rm log(O/H)_{\rm gas}} <9.4$,
with median value of 7.9. This quantity is directly linked to a photoionization model 
of a given interstellar metallicity, $Z$, and dust-to-metal mass ratio,
$\xi_{\rm d}$ (i.e., the fraction of metals depleted onto dust grains), as 
outlined in Sect. 2.3.2 of \citet{Gutkin16}. For reference, the solar interstellar
(dust+gas phase) and gas-phase oxygen abundances of the models are $12+{\rm log(O/H)}$=8.83
and $12+{\rm log(O/H)_{gas}}$=8.68 for $Z=Z_{\odot}=0.01524$ and
$\xi_{\rm d}=\xi_{\rm d \odot}=0.36$ respectively. The implications of these results are discussed
in the next section.

\section{Discussion}\label{sec:discussion}

\subsection{Nebular properties of narrow-line AGN at z=1.5-3}\label{sec:discussion_nebular}
By exploiting predictions from photoionization models, we investigated the excitation
properties of the ionized gas in the NLR of \civ-selected \tyii~AGN.  
Several works have performed similar studies on other samples, even though those samples
are more limited in terms of number of sources and line detections. 
The line ratios measured from the spectra of our \civ-selected \tyii~AGN, shown in
Fig.~\ref{Fig5}, are similar to those measured from the spectra of {\it z}$\sim$2 radio galaxies
and X-ray selected narrow-line QSO \citep{DeBreuck00,Vernet01,Szokoly04,Nagao06,Dors14}.
We review and discuss the main results from our analysis below.

{\it \textup{Ionization parameter.} }
Our \civ-selected AGN favor models with a relatively high volume-averaged ionization parameter
($-2.8 < $\, \logU \,$< -0.65$, Sect.~\ref{sec:line fitting}), but this is in line with those observed
in other samples of active galaxies. \citet{Nagao06} and \citet{Dors14} found
$-2.2<{\rm log} (U_{0})<-1.4$ and $-2.5<{\rm log}(U_{0})<-1.0$, respectively.
We note that the models of \citet{Nagao06} and \citet{Dors14} are parameterized in terms of
the ionization parameter computed at the inner edge of the gas distribution $r_{\rm in}$ ($U_{0}$=$U_{(r_{\rm in})}$),
the values of which are not directly comparable with the volume-averaged ionization parameter
considered here.  
For comparison purposes, we computed the ionization parameter from the \ciii/\civ\ ratio using
Eq.~1 of \citet{Dors14} for the 62 sources of our sample in whose spectra the \ciii\ line is detected. We obtained ionization parameters at the inner radius of the NLR in the range
$-2.2<{\rm log}(U_{0})<-1.0$. These values are similar to those found in previous works
and further validate our \civ-based criteria to select \tyii~AGN.

{\it \textup{Interstellar metallicity.} }
As discussed in Sect.~\ref{sec:models}, high super-solar metallicities, up to $\sim 6\,Z_{\odot}$
\citep[e.g.,][]{HamannFerland93, ShemmerNetzer02,Nagao06}, have been invoked to explain enhanced \nv\
emission in both BLR and NLR of AGN.
In our sample of 1.5$\,\leq\,${\it z}$\,\leq\,$3.0 \tyii~AGN  we estimated a subsolar or close
to solar interstellar (gas+dust phase) metallicity for 89\% of our objects.
In particular, only 5 of the 54 \civ-selected sources at $z>$1.9,
the redshift at which the \nv\ line enters the VIMOS spectral coverage, have supersolar
metallicity. This result is at odds with results presented in literature and probably
arises because we measured a larger number of different emission lines and adopted a wider range
of free parameters in our model grid (i.e., inner radius and internal microturbulence), which can explain the enhanced \nv\ emission observed in AGN spectra at $z\,\sim\,$2
without invoking high supersolar metallicities. 

\citet{Dors14} calibrated the C43 index (see Introduction) to derive the AGN metallicity. This
quantity is based on the strong \ciii, \civ,\ and \heii\ emission lines that are observed in the spectra
of active galaxies and does not consider the \nv\ lines. 
For the 57 sources for which the \ciii, \civ,\ and \heii\ lines 
are simultaneously present in the spectral range (excluding upper limits), 
we computed the C43 index and compared it with the metallicities inferred from our fitting. 
Specifically, we used the "upper branch" coefficients tabulated in Table 3 of \citet{Dors14}. 
We find that the C43 indicator yields a metallicity in the range
0.1$\,\lesssim\,$$Z/Z_{\odot}\,$$\,\lesssim\,$3.0, with $Z/Z_{\odot}\,>\,$2.5 for only 7 of 57 galaxies.
We note that the values for the metal abundances and solar metallicity adopted by \citet{Dors14}
slightly differ from those of this work. However, despite these different assumptions, which
prevent a real quantitative comparison, the ranges of Z/Z$_{\odot}$ obtained with the line
fitting and the C43 method agree overall.
The advantage of our model-dependent metallicity estimate is that it employs all of
the available nebular lines rather than relying on a small subset of them, as is the case for C43.

{\it \textup{Inner radius of the NLR.}}
As summarized in Sect.~\ref{sec:line fitting}, the majority of the line fluxes measured in
our sample favor models with $r_{\rm in}=90 \times (L_{\rm 45})^{0.5}\, {\rm pc}$,
where $L_{45}$ is the AGN luminosity, $L_{\rm AGN}$, expressed in units of $10^{45}\, {\rm erg\, s^{-1}}$.
A similar relation has also been found by \cite{Mor09}. A smaller inner radius for a given
AGN luminosity implies higher radiation pressure at the inner edge of the cloud (Sect. \ref{sec:photomodels},
see also \citealt{Dempsey18}).
This can be interpreted as UV lines originating from regions
closer to the central ionizing source, where the radiation pressure is higher for a radiation
field of a given intensity. This result confirms the wide range of ionization levels expected
in the AGN NLR, as was also shown by some studies of the rest-optical lines of \tyii~AGN
\citep[e.g.,][]{Richardson14}. 

{\it \textup{Internal microturbulence.}}
The majority of the rest-frame UV spectra of our \civ-selected \tyii~AGN are better explained
by models with internally microturbulent gas. To mimic the internal microturbulence and
dissipative heating, we followed \citet{BottorffFerland02} (see also \citealt{Kraemer07})\footnote{Specifically, 
we used the \texttt{Turbulence} and \texttt{Heat} command of the
photoionization code CLOUDY for the internal microturbulence and dissipative heating, respectively.}.

Our calculations were computed for a gas density of $10^2 < n_{\rm H} < 10^4$ cm$^{-3}$, that
is, a lower density regime than was used in the BLR models of \citet{BottorffFerland02}
and lower than the $n_{\rm H}=10^5$ cm$^{-3}$ of the NLR in \citet{Kraemer07}. 
This implies that the dissipative heating has a small effect on the lines of interest here,
whereas a major role in strengthening the intensities of high-ionization lines is played
by photoexcitation.  

The number of continuum photons that can be absorbed by the gas before it becomes 
optically thick to the incident radiation is higher when the gas is internally microturbulent.
These broader absorption profiles will affect resonant transitions, such as \civ\ and \nv, which
increases the contribution from photoexcitation. This is shown
in Fig.~\ref{Fig5}, where models with internal microturbulence have higher \civ/\heii \ and
\nv/\heii. It is worth nothing that a smaller inner radius and internal
microturbulence affect the line ratios in a similar way, and a combination of the two  
($r_{\rm in}=90~{\rm pc}$ and $v_{\rm micr}=100$~\kms) 
provides the highest intensities for the high-ionization \civ\ and \nv\ lines. 
We also note that more \hbox{sophisticated methods} for modeling the gas turbulence exist, such
as the use of hydrodynamical simulations \citep{Gray17}. To reproduce the complex physical
mechanisms associated with turbulent gas is beyond the goal of this work.

We relied on NLR properties inferred from photoionization models
and rest-UV emission lines. More reliable estimates of metal abundances can be obtained by
adding rest-optical lines (such as \oiii $\lambda5007$,\oiii$\lambda4959$,\oiii$\lambda4363$,
\oii$3727,3729,$ and the strong hydrogen recombination lines), from which we could have
direct measurements of the oxygen abundance. When models based on both the rest-UV
and rest-optical emissions are combined, high-to-low ionization states
of the NLR gas can also be investigated, and it can be studied how these different ionization levels correlate with the
radial extent of the narrow-line emitting region. Our sample therefore represents the ideal
target for near-IR spectroscopic observations, which are aimed at complementing our set of rest-UV
emission lines with the rest-optical lines. A comparative study of metal abundances
that are separately derived from rest-optical and rest-UV observations will be particularly relevant
in the context of spectroscopy of high-z AGN, for which we have to rely mainly on strong UV
lines. 

   \begin{figure}
   \centering
   \includegraphics[width=9cm]{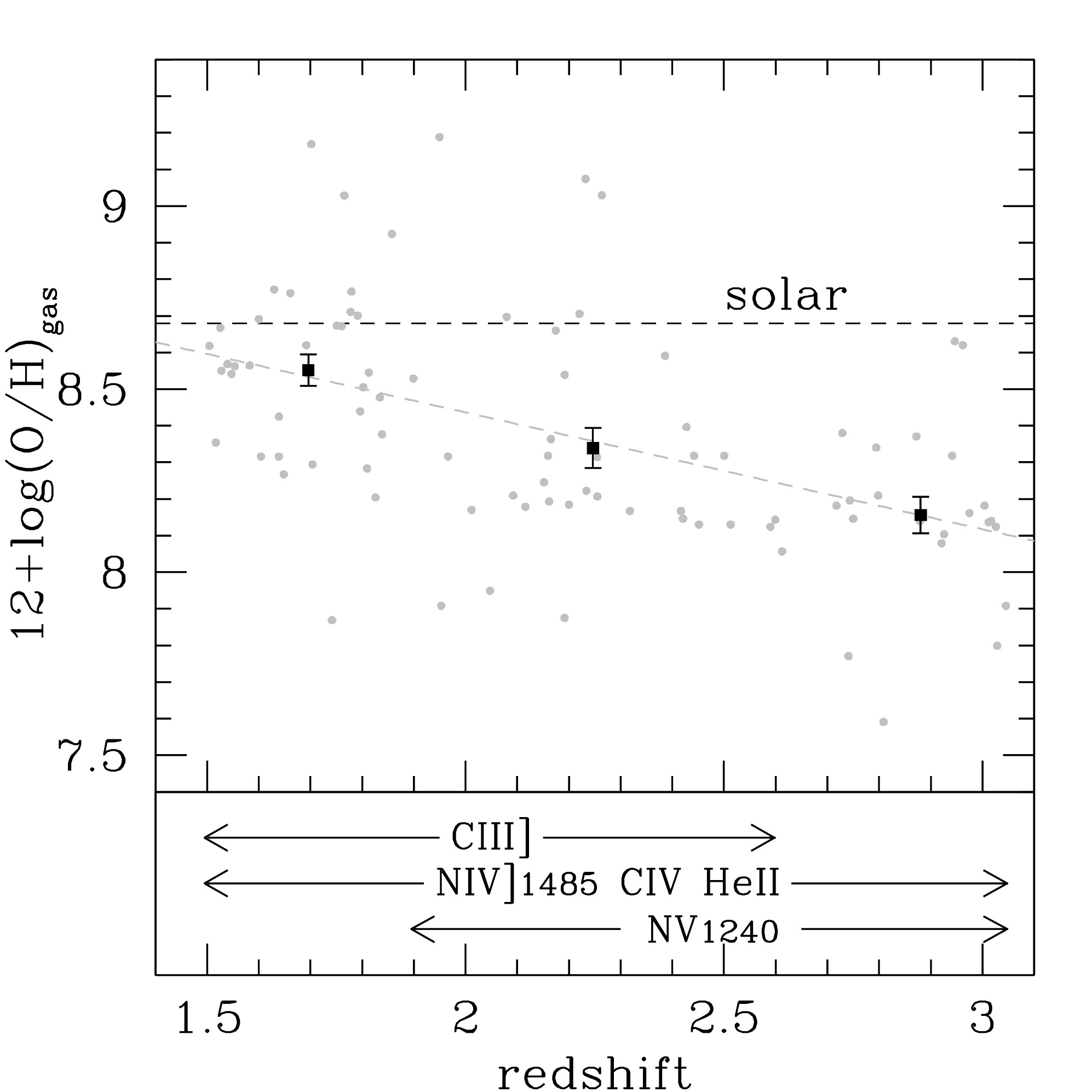}
   \caption{Evolution of the NLR gas-phase metallicity with redshift. The gray dots represent the 
   individual \civ-selected \tyii~AGN, while the black squares are the average oxygen abundance in
   three redshift bins. In the lower panel, the arrows highlight the redshift intervals where
   the quoted emission lines are visible in the observed spectral range.}
              \label{Fig6}
    \end{figure}

\subsection{Evolution with redshift of the NLR metallicity}\label{sec:discussion_zevo}

We investigated the chemical evolution of the NLR gas with redshift, and the
results are summarized in Fig.~\ref{Fig6}, where the gas-phase metallicity, expressed in terms of gas-phase oxygen abundance 12+log(O/H)$_{\rm gas}$ , of the \civ-selected
\tyii~AGN is plotted against redshift. Despite the large scatter, our measurements
indicate a moderate evolution of gas-phase metal content with redshift. A least-squares fit
to the data (the dashed straight line in Fig.~\ref{Fig6}) yields a decrement in the oxygen
abundance of $\approx$0.3 dex per unity of redshift. 
For the sample of 90 \tyii~AGN, the Spearman rank correlation coefficient is \hbox{$\rho$=-0.55,}
rejecting the hypothesis that no correlation is present at $>$~99.9\% confidence.
We also divided our sample into three redshift bins ({\it z}$\,\leq\,$1.9, 1.9$\,<\,${\it z}$\,\leq\,$2.6
and {\it z}$\,>$\,2.6) that correspond to different sets of emission lines used in the
comparisons with the models\footnote{At $z\,\leq\,$1.9 the \nv\,$\lambda 1240$ is still too blue
for the spectral range, while at $z\,>$\,2.6 the \ciii\,$\lambda 1909$ line moves out of 
the observed spectral range (see bottom panel of Fig.~\ref{Fig6}).} and contain 34, 33, and 23 objects,
respectively. We computed the average oxygen abundance in these three redshift bins
(black squares in Fig.~\ref{Fig6}) and obtained averaged values of 
$<$12+log(OH)$>$=(8.55$\pm$0.04, 8.34$\pm$0.06, 8.16$\pm$0.04).

The decline in gas-phase oxygen abundance with increasing redshift in the NLR of our \tyii~AGN is
further enforced by the fact that the redshift bins in which the intense \nv\,$\lambda 1240$ line is 
present are those with the lowest average metallicity, while an enhanced metallicity is needed to 
reproduce this line with standard photoionization models (see Sect.~\ref{sec:models}). Moreover,
the rate of the oxygen abundance decline that we measured is in good agreement with the evolution 
of the metallicity reported by \citet{Maio08} for a sample of high-$z$ star-forming galaxies: 
using their equation (2) and the average stellar mass of our sample ($2\times 10^{10}{\rm M}_\odot$),
we computed a decrement in the oxygen abundance of 0.285 per unity of 
redshift in the interval between 2.2$<\,${\it z}$\,<$3.5. This redshift range is slightly higher than but 
overlaps with the range considered in this work. A similar decline (0.3~dex offset
between {\it z}$\sim$2 and {\it z}$\sim$3.3) was observed by \citet{Onodera16}
for a sample of 41 normal star-forming galaxies. Observations of damped \Lya\ absorbers (DLAs)
at 0$<\,${\it z}$\,<$4.5 indicate that their metallicity decreases at a rate of $\sim$0.26 dex per unit 
redshift \citep{Prochaska03}. This rate of metallicity growth observed in DLAs is similar to the rate inferred
from the observed evolution of the mass–metallicity relation \citep[e.g.,][]{Erb06, Henry13}.

The SFR in galaxies ensures the production of heavy elements, whose local and global abundances in 
galaxies are modified by feedback processes, along with galactic winds, gas inflows, and accretion.
In general, the decrease in metal content of galaxies with increasing redshift is globally
accounted for by chemical evolultion models \citep[e.g.,][]{Dalcanton07,Calura09,Lilly13}, reproduced by semianalytical models 
and cosmological hydrodynamical simulations \citep[e.g.,][]{DeLucia04,Ma16,Dave11,Dave17,Torrey17},
and observed from {\it z}$\sim0$ to $\sim3$ in star-forming galaxies \citep[e.g.,][]{Maio08,Mannucci10,Yuan13}. 
There are many reasons for this observed decrease. First of all, high-z sources are much younger
and consequently less chemically evolved than present-day galaxies. In addition, higher z sources may
still be accreting a significant fraction of pristine gas and may host
more powerful outflows that can drive the metals outside the galaxy \cite[e.g.,][]{Yuan13}.

To our knowledge, this is the first time that a trustworthy metallicity evolution with redshift is
observed in the NLR gas of high-z \tyii~AGN. Previous studies on rest-UV line ratios of \tyii~AGN
\citep[e.g.,][]{Nagao06,Matsuoka09,Dors14} did not show a significant evolution with redshift.
The differences between our and previous results might depend on two main factors: first, our sample
was extracted from one single spectroscopic survey with a uniform selection technique, while previous works were based on a mixed collection of local AGN, radio galaxies, and X-ray selected
\tyii~AGN \citep[see][]{Nagao06}. Moreover, the fact that we observe a decrease in metallicity
with increasing redshift is most likely related to the different approach in inferring the NLR gas
metal content, exploiting more complex photoionization models (i.e., including the microturbulence) to
interpret multiple rest-UV emission lines rather than using single-line ratios.

\subsection{Mass-metallicity relationship in our type 2 AGN sample}\label{sec:discussion_MZR}
\begin{figure*}
   \centering
   \includegraphics[width=18cm]{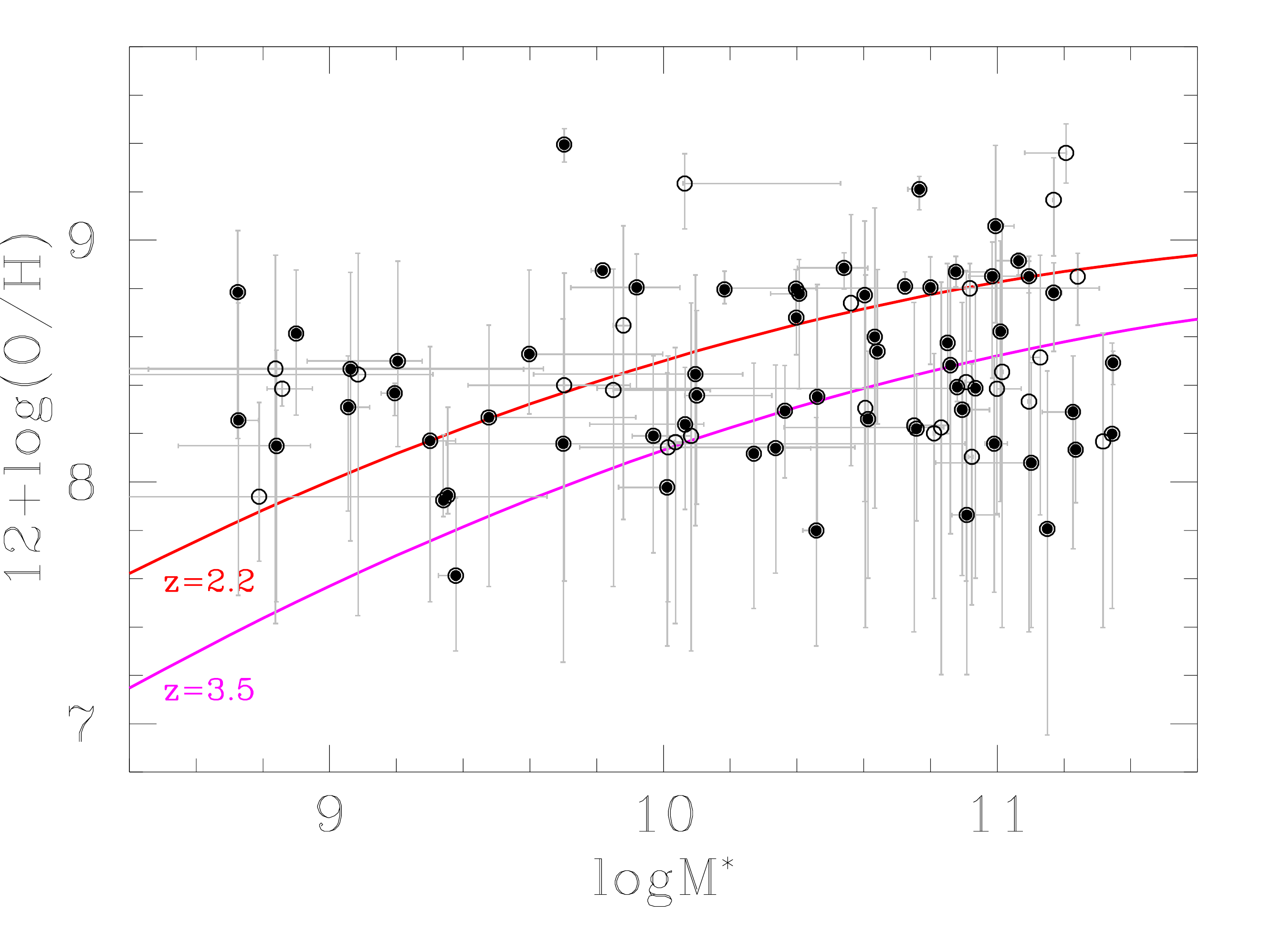}
   \caption{Mass-metallicity relation for the \civ-selected \tyii~AGN. The filled
   symbols mark the 61 objects with reliable measurements (emission lines detected 
   with S/N$>$3 and the SED fit with 0.5$\,< \chi^2_{reduced} <\,$5). 
   The solid lines show the MZ relation at {\it z}=2.2 and 3.5 from equation (2) of 
   \citet{Maio08}.}
              \label{Fig7}%
    \end{figure*}

After computing both the host galaxy stellar masses and NLR gas-phase metallicities,  
we investigated whether the mass-metallicity relationship (MZR) also held for our sample of \tyii~AGN.
Figure~\ref{Fig7} shows the mass-metallicity data for 88 \civ-selected \tyii \ AGN for which reliable 
measurements are available\footnote{Two \tyii~AGN do not have a good measurement of the 
host stellar mass, see Sect.~\ref{sec:masses}.} along with the MZR at {\it z}=2.2
(red curve) and {\it z}=3.5 (purple curve) from equation (2) of \citet{Maio08}. 
The vertical error bars in the figure represent the 1$\sigma$ (68\%) uncertainties resulting from
the 16th and 84th percentiles of the posterior distribution function of the gas-phase metallicity,
while the horizontal lines indicate the 1$\sigma$ errors on the stellar masses computed following 
\citet{Bongiorno12}\footnote{the error bars cover the range of values corresponding to the solutions
for which $\chi^2 = \chi^2(sol) - \chi^2(best) \leq 1.0$}.

Although our gas-phase metallicities are in the same range, at least for log(M$_*$) higher than 9.5,
as those computed by \citet{Maio08} for star-forming galaxies at similar redshifts, we do not find a
statistically significant MZR in our data. For the full sample of \civ-selected \tyii~AGN we find in a least-squares fitting of the data a shallow slope $\alpha$=0.08 of the relationship, and the
corresponding Spearman rank correlation coefficient cannot reject with high significance the 
null hypothesis of no correlation.
In order to understand whether the estimates of both stellar mass and gas metallicity in the 
whole \tyii~AGN sample can suffer problems due to either insufficiently precise measurements
or to host galaxy contamination, we repeated the least-squares fitting to the possible MZR
and its statistical evaluation in different subsamples. In all the cases the relationship is shallow
and less statistically significant than in the full sample. In Table~\ref{table:lsqfit} we show
the slope and r.m.s. of the least-squares fitting along with the correlation coefficient for 
1) the global sample; 2) the subsample with good measurements (all the emission lines detected with S/N$>$3
and the SED fit with 0.5$\,< \chi^2_{reduced} <\,$5); 3) the \tyii~AGN detected
in the X-rays with L$_{\rm 2-10keV}$$>$$6\times10^{42}erg\,s^{-1}$; 4) the sample of \tyii~AGN
with intense \civ \ emission-line (EW$_0>$15\AA); 5) the sample of \tyii~AGN with relatively
weak \civ \ emission-line (EW$_0<$15\AA); 6) the sample of \tyii~AGN with a photometry that
requires a SED fitting with an AGN dominance$>$1; and finally, 7) the sample of AGN with a relevant
galaxy component in the photometry, that is, AGN dominance $<$1. The values listed
in Table~\ref{table:modelpar} show that the mass-metallicity relationship is always extremely flat and
statistically compatible with no correlation between the two quantities.
Therefore, we can safely conclude that at least for our \civ-selected type 2 AGN sample, a strong
relationship is not observed between the NLR gas metallicity and the stellar mass of the host galaxy.

   \begin{table}
      \caption[]{MZR in different \civ-selected AGN samples}
         \label{table:lsqfit}
     \resizebox{0.45\textwidth}{!} {         
       \begin{tabular}{lcccc}
            \hline
            \noalign{\smallskip}
            sample & N$_{\rm OBJ}$ & $\alpha^{(1)}$ & r.m.s.$^{(1)}$ & $\rho^{(2)}$ \\
            
            \hline
            \noalign{\smallskip}
          all & 88 & 0.08 & 0.31 & 0.16 \\
          good measur. & 61 & 0.06 & 0.31 & 0.14 \\
          X-ray & 49 & 0.05 & 0.31 & 0.06 \\
          high EW(\civ) & 56 & 0.05 & 0.27 & 0.04 \\
          low EW(\civ) & 32 & 0.17 & 0.36 & 0.20 \\
          AGN domin. & 61 & 0.04 & 0.28 & 0.07 \\
          no AGN domin. & 27 & 0.09 & 0.34 & 0.01 \\  
            \noalign{\smallskip}
            \hline
         \end{tabular}
  }
  
  \tablefoot{$^{(1)}$ slope and r.m.s. from the least-squares fitting of the samples. 
  $^{(2)}$ Spearman rank correlation coefficient.}

   \end{table}

The absence of the MZR could be ascribed to sample selection effects. By selecting
powerful high-redshift AGN with intense emission lines, such as the \civ, we culled host
galaxies that may not be representative of the global star-forming galaxy population at these
redshifts.
The distribution of the host galaxy masses shown in Fig.~\ref{Fig4} already clearly shows that the distribution is strongly biased toward the tail with the highest stellar masses. 
Moreover, our selection required a strong high-ionization emission line in the UV 
range, which might have biased our sample toward objects with a strong ionization continuum, 
with a massive central black hole, and with mild-to-absent galactic extinction. 
All these properties may be connected more with an 
evolved and massive host galaxy bulge than with a late-type galaxy rich in dust and gas.

On the other hand,  Fig.~\ref{Fig7} shows host galaxies that even if they 
have stellar masses lower than $1.5\times 10^9\,M_{\odot}$, show an NLR gas metallicity of
12+log(OH)$_{\rm gas}>8$, and such values are expected for more massive galaxies. These
objects, which keep the MZR relatively flat within the mass range explored by our sample,
cannot be simply explained as due to selection effects. The MZR for star-forming galaxies
has been interpreted as a consequence of the interplay between star formation efficiency, 
inflows of metal poor-gas, gas outflows due to galactic winds, and consequent metal depletion
\citep[e.g.,][]{Mannucci10}. Moreover, lower mass galaxies are more metal poor because their
potential well is smaller \citep{Dayal13}. The presence of an AGN could strengthen the potential well in the
central part of the NLR, where most of the high-ionization emission lines are emitted 
\citep[in particular the UV lines;][]{Costantini16}, reducing the metal depletion in the NLR
gas across the range of host galaxy stellar masses. Another possible explanation of the observed lack
of relation between the gas-phase metallicity of the NLR and the mass of the host is that 
strong AGN-driven outflows can carry metal-rich gas from the BLR to radial distances of up to few kiloparsec.
It is well known that the metallicity of the BLR is higher than that of the host galaxy \citep{Xu18, Thomas19}
and does not seem to evolve with redshift \citep{Warner03,Nagao06}.
Even if the metal-rich gas in the NLR can be supplied by massive stars from the bulge, it could also
be polluted by metals delivered by outflows emerging from the central regions of the AGN,
which enhance the metallicity in the NLR and weaken the relationship with the stellar mass content
of the host galaxy. A connection between the two AGN regions has also been suggested
by \citet{DuP14}, who found a strong correlation between NLR and BLR metallicities for 31 low-z \tyi~AGN.

The lack of MZR in our sample is at odds with the relation recently found by \citet{Matsuoka18}
in high-z type-2 AGN.
There can be many reasons for this discrepancy, and they include differences in terms of
sample selection, mass and redshift ranges of the objects, and photoionization models. 
Our sample is uniformly selected and measured from the zCOSMOS-deep galaxy sample, 
while \citet{Matsuoka18} collected mixed spectra of previously confirmed AGN through either radio
or X-ray selection. Our \civ-selected \tyii\ AGN sample comprises 88 objects in a smaller
redshift range compared to the 1.2$\,\leq z \leq\,$4.0 range of the 28 objects of the previous study. 
Moreover, the stellar masses of about 40\% (35/88) of the \civ-selected AGN are lower than then
minimum value of \citet{Matsuoka18}, that is, log$(M_{\star}) = 10.2$ M$_{\odot}$.
Finally, the suite of photo-ionization models used in this work differs from that of 
\citet{Matsuoka18} in terms of the values of gas densities, the set of metal
abundances, and the inclusion of dust physics and metal depletion. 
Because the two setups of the photoionization models are so different, it is difficult
to perform a quantitative comparison between the metallicity values inferred with the two model grids. 
\section{Summary and conclusions}\label{sec:summary}
\begin{itemize}
\item
This paper presented a sample of 192 \civ-selected AGN in the redshift range 1.45$<${\it z}$<$3.05, 
extracted from the zCOSMOS-deep survey. Ninety of them are classified as \tyii~AGN on the basis of the
FWHM of the emission lines (Sect. \ref{sec:AGNclassification}).
\item
The average spectral properties in the rest-frame UV range are compatible with other (brighter)
optical AGN samples. About three-quarters of the sample are detected in X-rays based on the Chandra COSMOS-Legacy
survey, but the detection rate is significantly different for the two AGN spectral classes:
more than 94\% of \tyi~AGN are detected by Chandra, while for the \tyii~AGN, the fraction decreases to 57\%,
meaning that they are likely obscured in the X-rays (Sect. \ref{sec:X-ray}). 
\item
We have accurately measured fluxes and widths of seven \hbox{UV} emission lines in the VIMOS spectra
of the \tyii~AGN, and we have exploited the large set of photometric data available in the COSMOS field
to estimate the stellar masses of their host galaxies using a two-component (AGN and galaxy) SED-fitting technique. 
The host galaxy stellar masses of the \tyii~AGN are in the range 8.5$\times$10$^8\,-\,$2.2$\times$10$^{11}$M$_\odot$; the distribution is skewed toward high masses and peaks at $\sim$10$^{11}$M$_\odot$~(Sect. \ref{sec:masses}).
\item
We find that the observed emission-line ratios involving the \civ\ and \nv\ emission lines 
can be best explained by models with internally microturbulent clouds and an inner radius
of the NLR gas distribution  of $r_{\rm in}=90 \times (L_{\rm 45})^{0.5}\, {\rm pc}$.
All but four ($\sim 95\%$) of the \tyii~AGN lie in the regions
of the UV-based emission-line ratio diagnostic diagrams occupied by these modified
AGN photoionization models. This further strengthens the proposed AGN selection through 
strong \civ\ emission (Sect. \ref{sec:diagnostics}).
\item
We have implemented a simple spectral fitting to estimate the oxygen abundance of the NLR gas in
our \civ-selected \tyii~AGN, finding subsolar or solar interstellar metallicity 
for 89\% of the sample (Sect. \ref{sec:line fitting}). 
At $z>1.9$, only 5 of 54 \tyii~AGN show super-solar metallicity, even though at this redshift
the \nv\ line enters the spectral range. We explained the observed strength of the \nv\ emission by reducing the inner radius of the NLR and by introducing microturbulent
clouds \citep[see also][for the latter]{Kraemer07}, without the need of invoking highly super-solar
metallicities (Sect. \ref{sec:discussion_nebular}).
\item
The gas-phase metallicity of the NLR in our \tyii~AGN exhibits a statistically significant evolution
with redshift in the range covered by our sample \hbox{(1.5$\,<${\it z}$\,<$3.0)}. We measure a decrement
in the oxygen abundance of $\approx$0.3 per unity of redshift. This value is comparable to other
estimates obtained in star-forming galaxies at similar redshifts (Sect. \ref{sec:discussion_zevo}).
\item
The mass-metallicity relationship is not observed in our \civ-selected sample of \tyii~AGN, 
even when differently selected subsamples were considered (Sect. \ref{sec:discussion_MZR}).
The lack of a statistically significant MZR could be ascribed to sample selection effects because the high-ionization \civ \ emission line can cause culling of massive and
dust-free hosts, which are not representative of the star-forming galaxy population at these 
redshifts. These results could also suggest that the NLR metallicity is not a good proxy
of that of the host galaxy because an AGN can affect the 
gas supply and metal depletion in these regions.

\end{itemize}

We have adopted single-density ionization bounded models of 
pure AGN nebular emission to interpret the observed excitation properties of the NLR emitting gas. 
Future works should focus on more complex and tailored spectral decompositions accounting, in addition
to the AGN component, for the contribution of star-formation to the observed SED. 
Moreover, the interpretation of the physical mechanisms responsible for the enhanced emission of the 
resonant \nv\ line in AGN will strongly benefit from advanced models such as those based on hydrodynamical
simulations of turbulent gas, as well as a proper treatment of the resonant lines by means of radiative
transfer through the neutral medium. This will have relevant implications for the detection and study 
of potential AGN host galaxies close to the epoch of reionization, as indicated by some tentative
detections at $z\gtsim 7$ that have been presented in the literature \citep[e.g.,][]{Laporte17,Mainali18}.
To conclude, combining the rest-UV emission of our sample with additional information from rest-optical
spectra from other facilities \citep[e.g. FMOS-COSMOS;][]{Silverman15} will allow us to better constrain the
models and compare the physical properties inferred from the rest-optical with those from rest-UV.
This is particularly important for future studies at higher redshifts with forthcoming facilities,
such as the {\itshape James Webb Space Telescope,} which will probe both the rest-frame and optical UV
emission of galaxies at $z \gtrsim 3,$ and {\itshape EUCLID}, which  will unveil the rest-UV emission of
the more luminous AGN at $z \gtrsim 6$.

The figures of the VIMOS spectra and the ACS images of the \civ-selected \tyii~AGN, along with figures
showing results from the SED fitting described in Sec. \ref{sec:masses} and spectral fitting described in
\ref{sec:line fitting}, are available electronically from \url{http://www.bo.astro.it/\~mignoli/CIV_zCOSMOS/ty2_table.html}.
The data is also available at \url{http://cdsarc.u-strasbg.fr}.

\begin{acknowledgements}
    We acknowledge financial support from the agreement ASI-INAF n.~2017-14-H.O.
    A.F. acknowledges support  from  the  ERC  via  an  Advanced
        Grant under grant agreement no. 321323-NEOGAL and no. 339659-MUSICOS.
        A.C. acknowledges the support from the grants PRIN-MIUR 2015 and ASI
        n.I/023/12/0 and ASI n.2018-23-HH.0.
        Y.P. acknowledges National Key R\&D Program of China Grant 2016YFA0400702,
        and NSFC Grant No.~11773001 and 11721303.
\end{acknowledgements}

%
%

\end{document}